\documentclass[12pt,preprint]{aastex}

\def\logz{\lbrack\hbox{M/H}\rbrack}
\def\feh{\lbrack\hbox{Fe/H}\rbrack}

\newcommand{\mean}[1]{\langle #1 \rangle}

\slugcomment{To appear in the Astronomical Journal}
\shorttitle{M31 Variables}
\shortauthors{Dolphin et al.}

\begin{document}

\title{Short-Period Variable Stars in the M31 Halo}

\author{Andrew E. Dolphin}
\affil{Steward Observatory, University of Arizona, Tucson, AZ 85721}
\email{adolphin@as.arizona.edu}

\author{A. Saha}
\affil{Kitt Peak National Observatory, National Optical Astronomy Observatories\linespread{1.0}\footnote{NOAO is operated by the Association of Universities for Research in Astronomy, Inc. (AURA) under cooperative agreement with the National Science Foundation.  The WIYN Observatory is a joint facility of the University of Wisconsin-Madison, Indiana University, Yale University, and the National Optical Astronomy Observatories.}, P.O. Box 26372, Tucson, AZ 85726}
\email{saha@noao.edu}

\author{Edward W. Olszewski}
\affil{Steward Observatory, University of Arizona, Tucson, AZ 85721}
\email{eolszewski@as.arizona.edu}

\author{Frank Thim}
\affil{Kitt Peak National Observatory, National Optical Astronomy Observatories$^1$, P.O. Box 26372, Tucson, AZ 85726}
\email{thim@noao.edu}

\author{Evan D. Skillman}
\affil{Astronomy Department, University of Minnesota, Minneapolis, MN 55455}
\email{skillman@astro.umn.edu}

\and

\author{J.S. Gallagher and John Hoessel}
\affil{University of Wisconsin, Dept. of Astronomy, 475 N. Charter St, Madison, WI 53706}
\email{jsg@astro.wisc.edu, hoessel@astro.wisc.edu}

\begin{abstract}
We report the findings of a new search for RR Lyraes in an M31 halo field located 40 arcminutes from the nucleus of the galaxy along the minor axis.  We detected 37 variable stars, of which 24 are classified as RR Lyraes and the others are ambiguous.  Estimating a completeness fraction of $\sim 24\%$, we calculate that there are approximately 100 RR Lyraes in the field, which is consistent with what is expected from deep HST color-magnitude diagrams.  We calculate a mean magnitude of $\mean{g} = 25.15 \pm 0.03$, which we interpret to mean that the mean metallicity of RR Lyraes is significantly lower than that of the M31 halo as a whole.  The presence of ancient, metal-poor stars opens the possibility that, initially, the M31 halo appeared much like the Milky Way halo.
\end{abstract}


\section{Introduction}

The Andromeda galaxy (M31) provides a unique opportunity to study the structure and evolution of massive spirals, as it is the closest such system that can be studied from outside (i.e., line of sight depth is not a major problem).  As such, a comparison to our Galaxy allows us to address the question of variety in the evolutionary histories of massive spirals.  For example, van den Bergh (2000, 2003) has suggested that the chemical evolution of M31 was much more rapid than that of the Galaxy, and this might point to the origin of M31 as the early merger of two (or more) relatively massive metal-rich ancestral objects.  Variable stars provide a unique and powerful tool to inspect the stellar populations and and star formation histories of nearby galaxies (Mateo 1998; 2000; Saha 1999), and here we employ observations of variable stars in the M31 halo to provide new constraints on the evolution of M31.

\citet{mou86} made the first study of the stellar populations in the M31 halo.  An important result of that study was that the M31 halo red giant branch is much redder than those of Galactic globular clusters.  This was interpreted as a sign that the mean metallicity was $\mean{\feh} = -0.6$, a result that has been confirmed by deeper imaging \citep{dur01}.  Shortly after this work, a second milestone study of the M31 halo was carried out by Pritchet \& van den Bergh (1987; hereafter PvdB), who imaged a field 40 arcminutes from the nucleus along the SE minor axis to search for variable stars.  They reported the discovery of 30 RR Lyraes, and assuming a completeness of 25\% estimated that the field contained 120 such objects.  Compared with the number of red giants, this implied that the frequency of RR Lyraes in the M31 halo is comparable to that found in RR Lyrae-rich globular clusters.  As noted in their work, this frequency was surprisingly high given that the \citet{mou86} metallicity was significantly higher than that measured in RR Lyrae-rich Galactic globular clusters.

The number density of RR Lyraes estimated by PvdB has been called into question by deep HST color-magnitude diagrams of the M31 halo, which show few stars in the region where one would expect RR Lyraes.  A recent ultra-deep ACS CMD \citep{bro03} is the best such example, finding roughly an order of magnitude fewer stars in that part of the CMD than would be expected from the PdvB RR Lyrae frequency.  Several possibilities exist for the discrepancy in the RR Lyrae frequency.   The pointings used by PvdB and \citet{bro03} were different, and findings of streams in the M31 halo \citep{iba01} raise the possibility of position-dependent populations.  A second possibility is that accurate photometry of stars at $B = 25.7$ was at or beyond the capabilities of CCDs in the mid 1980s, and perhaps the earlier RR Lyrae experiment detected specious objects in addition to bona fide RR Lyraes.

To address this discrepancy with a straightforward approach, we repeated the PvdB RR Lyrae experiment and present our findings in this paper.  Section \ref{sec_obs} describes our data and reduction procedures, and the following sections describe the number of RR Lyraes and their properties.

\section{Observations and Reduction \label{sec_obs}}

We made repeated observations of a field in the M31 halo over the four nights 16$-$19 November 2001 with the MIMO camera (whose performance is described and characterized by Saha et al. 2000) on the 3.5m WIYN telescope.  Our field is located about 40' along the southeast minor axis of M31, and contains part of the field of \citet{mou86} and all of the field of PvdB.  About half of night two and most of night three was lost due to weather; on the first and fourth nights we were able to observe M31 for approximately 8 hours per night.   In total, we had 35 usable images (28 $g$ and 6 $r$) over a 3.3 day baseline, 33 of which were 1800 seconds and the others were slightly shorter due to telescope and instrument operation modalities.  Our log of observations is listed in Table \ref{tab_obslog}.
\placetable{tab_obslog}

Photometry was obtained using a modified version of HSTphot \citep{dol00}, a PSF-fitting photometry package.  The original package was designed solely for use with WFPC2 images; the modified version used here is for general use.

The photometry process involved several steps.  First the creation of a reference image, which would be used to measure star positions and determine a standard coordinate system.  We used the best-seeing images available for this purpose: n1071, n1072, n1074, and n1075 in $g$, and n1052, n1073, and n2035 in $r$.  The $g$ reference image is shown in Figure \ref{fig_image}.  The upper-right portion of chip two was masked out because a probe was vignetting the camera during those observations; unfortunately we had no other suitable observations to fill in that gap in our reference image.
\placefigure{fig_image}

After this, photometry was run on the individual images, using star positions determined from the deep photometry obtained in the same filter.  The use of known star positions made it possible to obtain accurate photometry in images of poorer quality, as well as ensuring that our list of stars would be the same at each epoch.

After completing this procedure, we found that the photometry appeared too shallow to detect the fainter RR Lyraes.  To compensate, we ran a second set of epoch photometry in which consecutive $g$ exposures were coadded when possible.  We were able to create 12 coadded images (four $g$ images could not be combined) and obtained deeper photometry.  The 16 epochs are listed in Table \ref{tab_obslog} under the ``Epoch'' heading.  While giving fewer data points per star, the $\sim 0.3$ magnitude increase in our depth was essential for a fairly complete recovery of RR Lyraes.

Calibrations were made from our night 1 data, which was our only photometric night of the run, using observations of several Gunn standard stars: BD +28 4211, BD +17 4708, and Ross 34 \citep{ken85}.  Using M31 images from similar airmass (n1050 and n1052), we created a set of secondary standard stars in the M31 halo field, to which all of our photometry was transformed.  Residuals in the night solutions were no greater than 0.015 magnitudes (rms).  Our deep CMD (from the reference images) is shown in Figure \ref{fig_cmd}.
\placefigure{fig_cmd}

To provide a check on our photometry and calibrations, we produced an independent calibration using different software and photometry using DoPHOT \citep{sch93}.  The agreement is better than 0.01 magnitudes at all brightness levels, indicating excellent agreement of both the calibrations and the PSF-fitting photometry.  We show a comparison of the two sets of photometry in Figure \ref{fig_photcomp}.
\placefigure{fig_photcomp}

\subsection{Image Subtraction Photometry}

Since we knew that any RR Lyraes in M31 would be at the detection limits, we made a second set of reductions of the $g$ images using a image subtraction technique that has been incorporated into the modified HSTphot.  The principles of this procedure are similar to those laid out by \citet{ala98}.  A reference image was constructed from the best-seeing epochs, and was subtracted from each individual image.  To accomplish the subtraction, one must align (rotate and shift) the reference image to match coordinates, convolve with a kernel to match PSFs, and scale the image to the same count levels.  The resulting difference image can then be photometered using PSF-fitting code.  A variable star will show up as a positive (if brighter than the reference image) or negative (if fainter) feature in the subtracted image.

We introduce one significant difference: rather than calculating the subtraction kernels directly from the images, we calculated them from the PSFs measured by the photometry program.  Making this change creates potential problems if the PSFs are incorrect, but allows us to ignore problems of bad pixels, background differences (a major issue here because of scattered light), and smoothing from interpolation.  Another minor change is that our routine does not assume the noise of the reference image to be zero.

The reference image was then convolved by this kernel, scaled, and subtracted from the epoch image.  We show part of the difference image for epoch 11 in Figure \ref{fig_res}, as well as the residual from our PSF-fitting photometry.  Comparing those, we see that stars are subtracted very well in both (which was expected since the PSF is well-measured and the field is not crowded), but that extended objects and stars below the detection limit are subtracted much more accurately using image subtraction.  This allows for more reliable measurements of variability, despite the fact that we are working well outside the crowded-field regime where image subtraction techniques are normally applied.  The only notable weakness is that the noise is increased by the subtraction process.
\placefigure{fig_res}

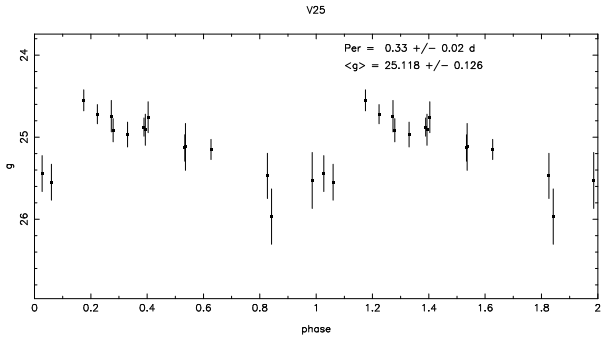
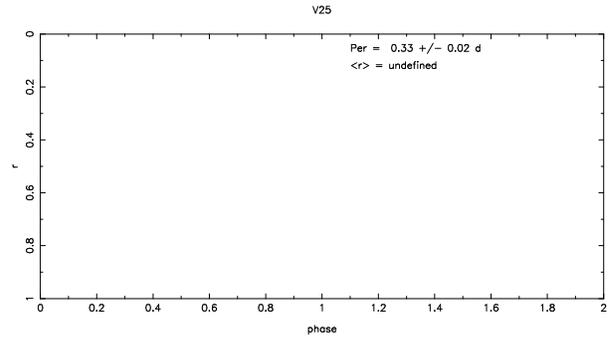
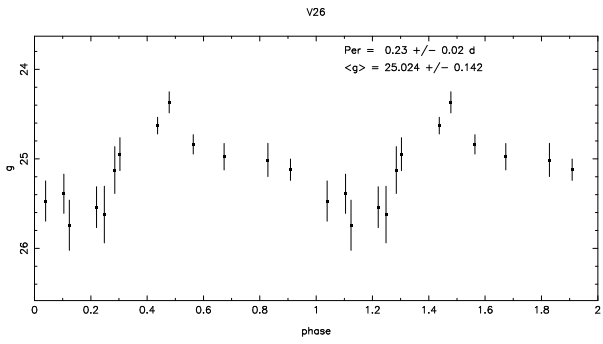
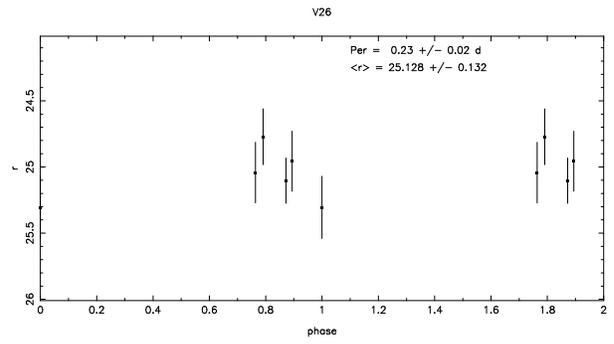
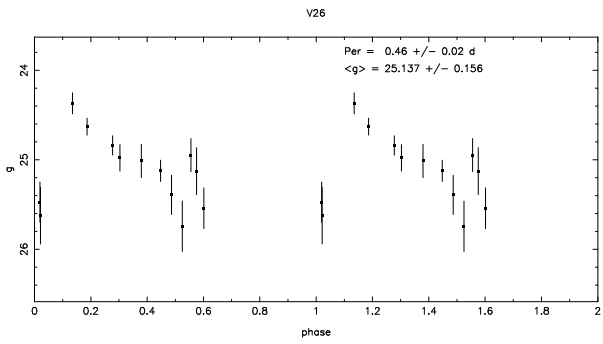
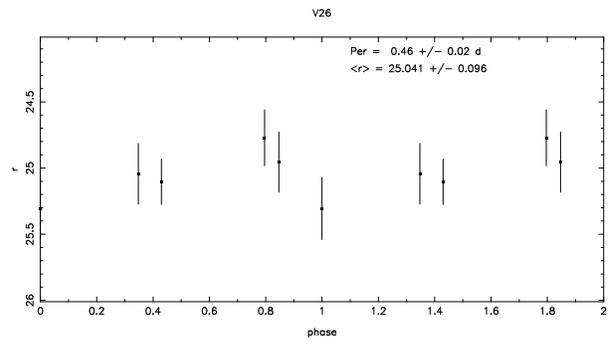
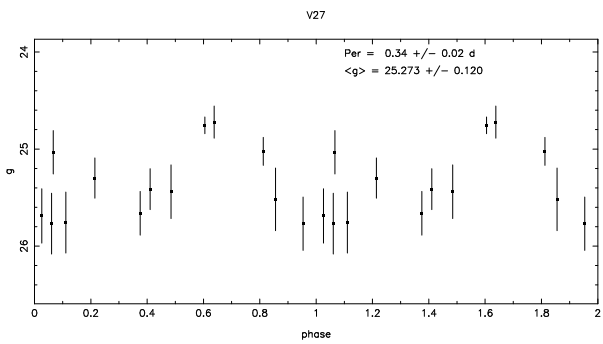
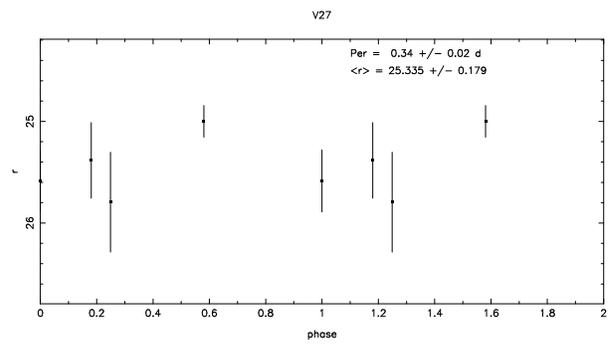
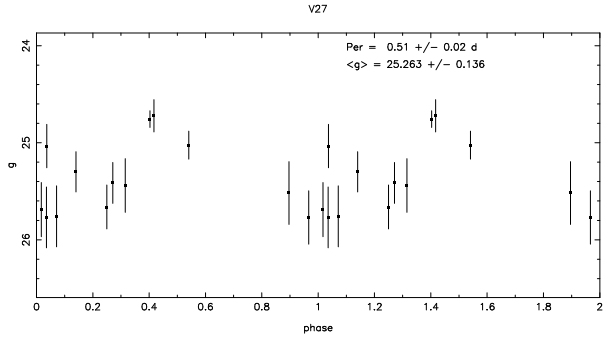
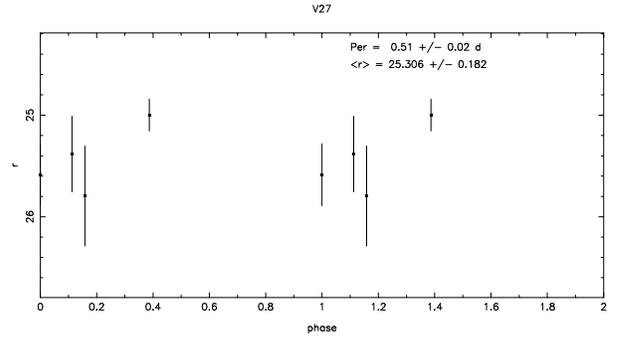
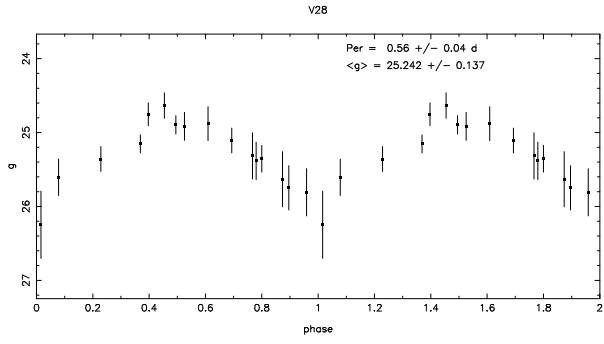
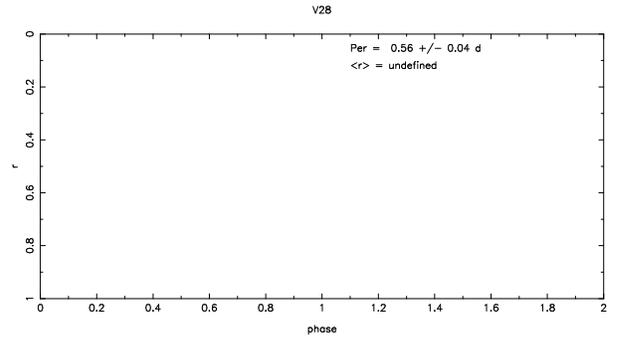
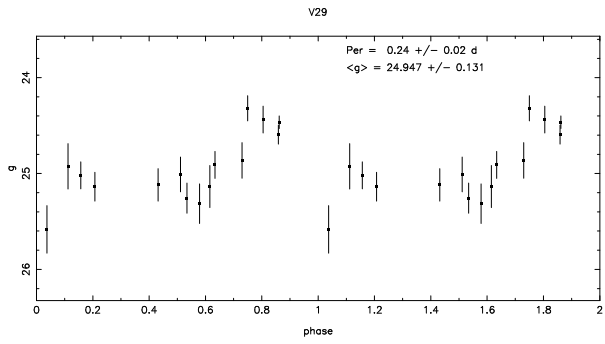
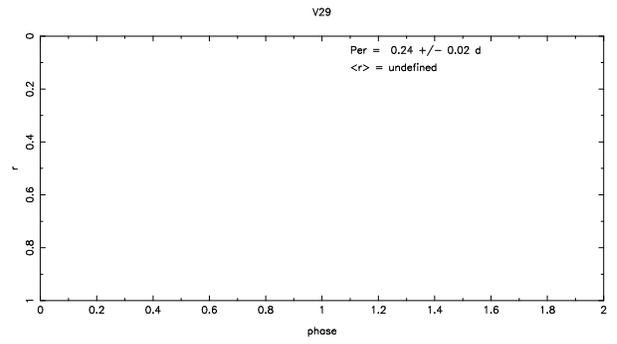
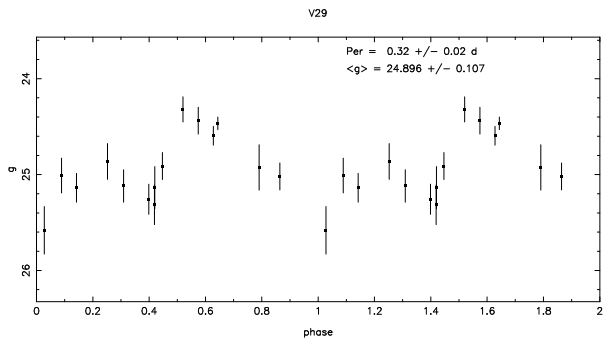
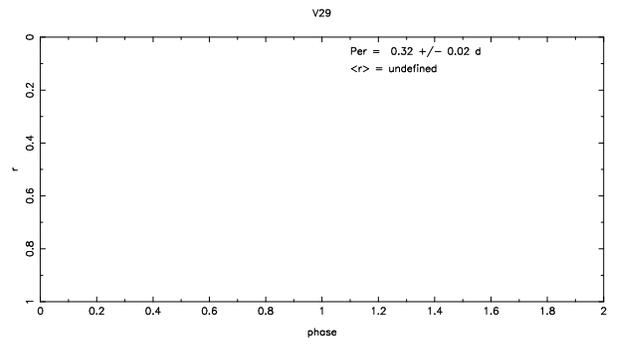
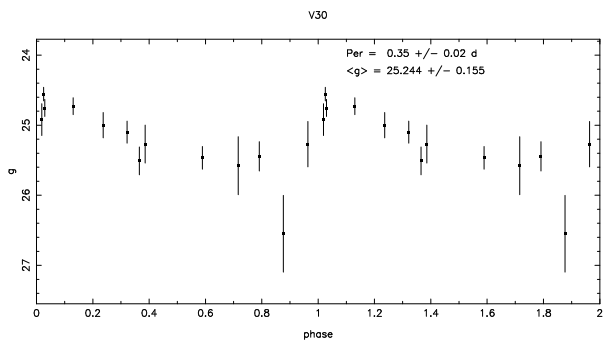
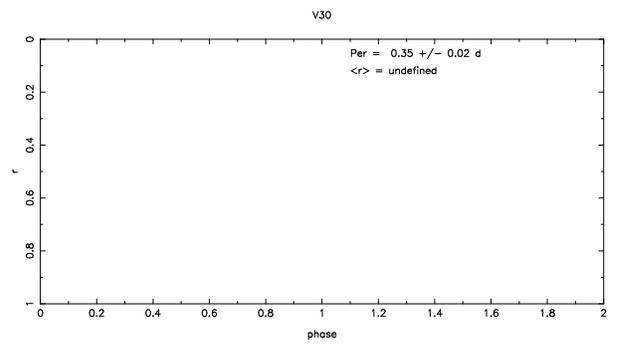
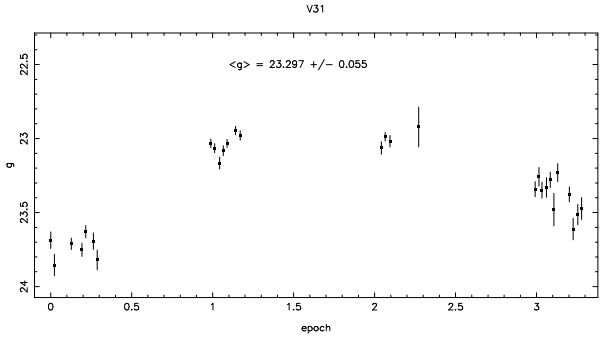
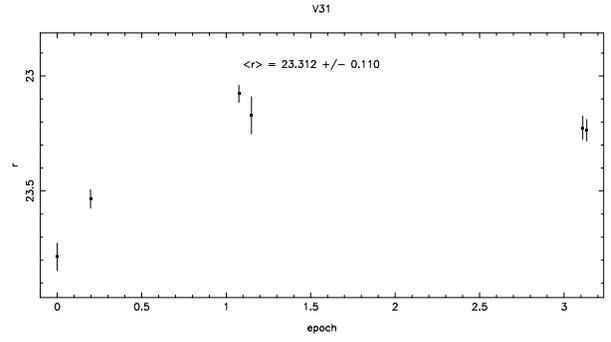
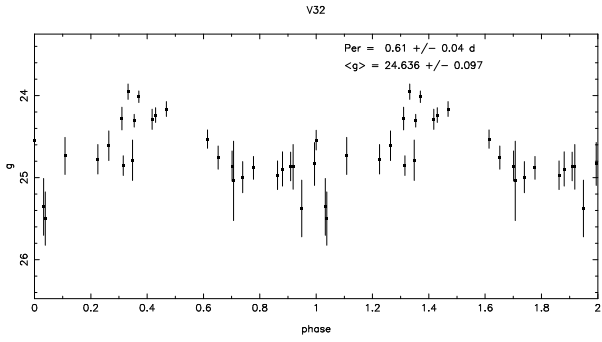
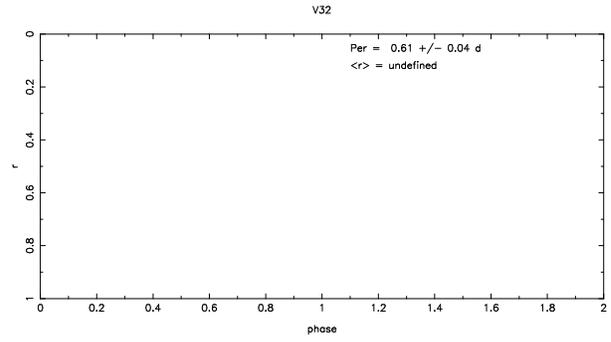
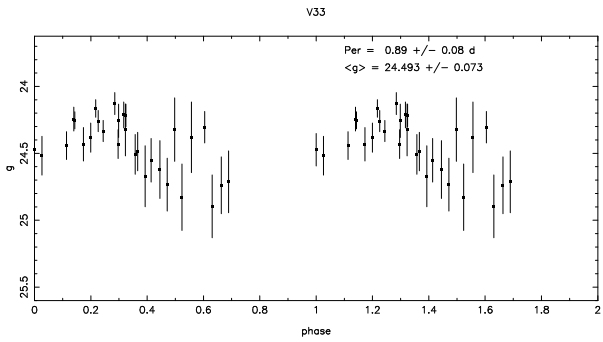
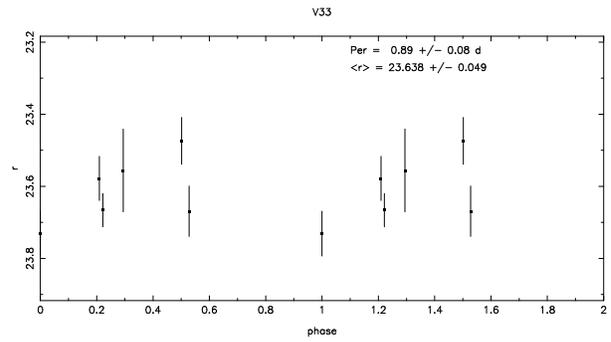
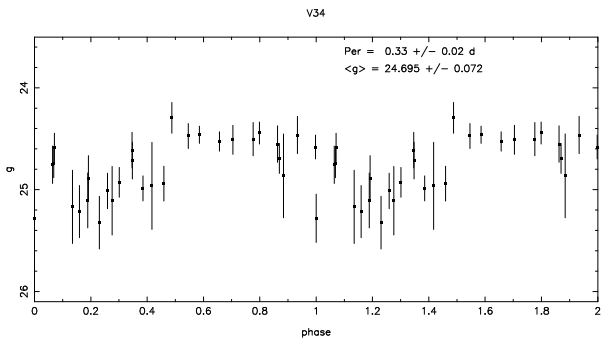
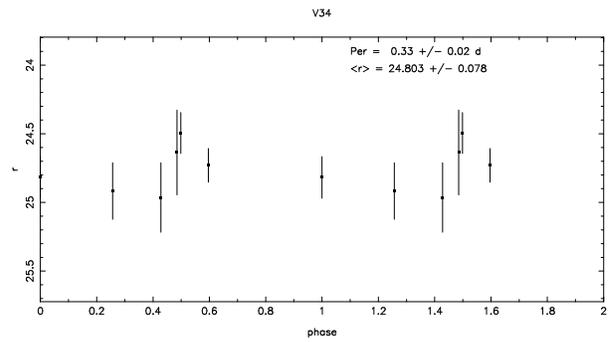
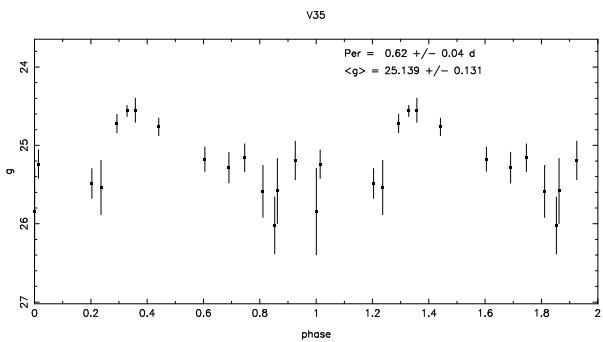
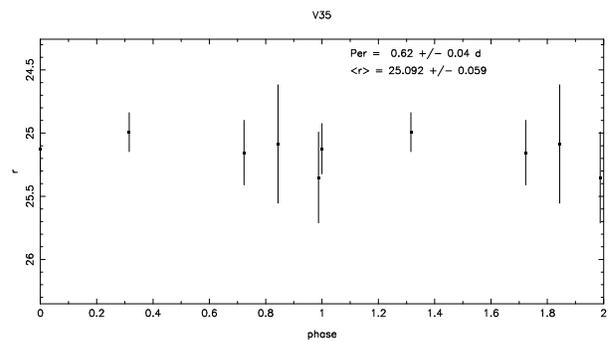
\begin{figure}
\plotone{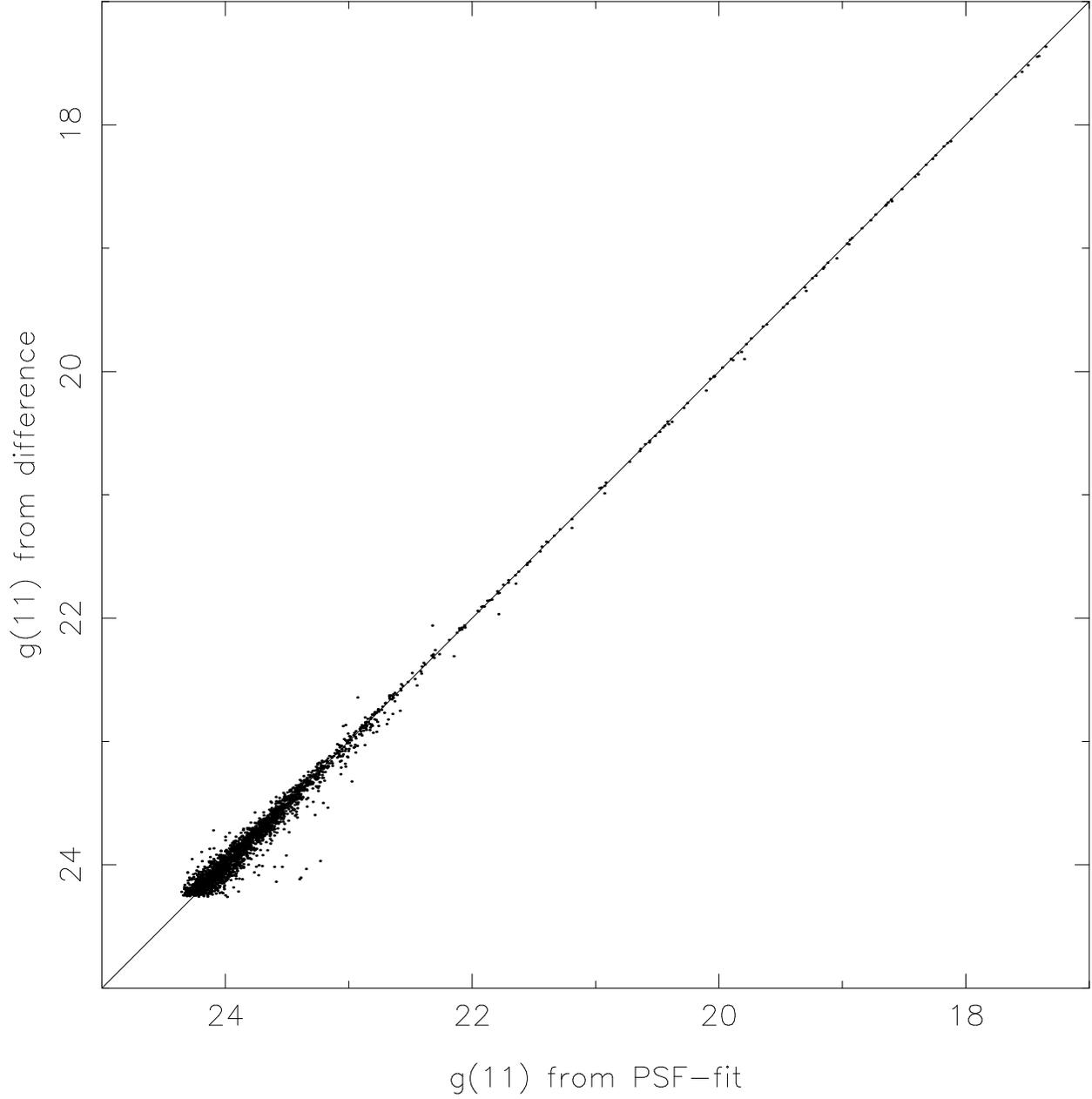}
\caption{A comparison of epoch g11 magnitudes from chip 1.  The x axis shows $g$ magnitudes from the PSF-fitting photometry, while the y axis shows $g$ magnitudes calculated by our image subtraction code.  The diagonal line is $y=x$, not a fit to the data.}
\label{fig_mags}
\end{figure}

\begin{figure}
\plotone{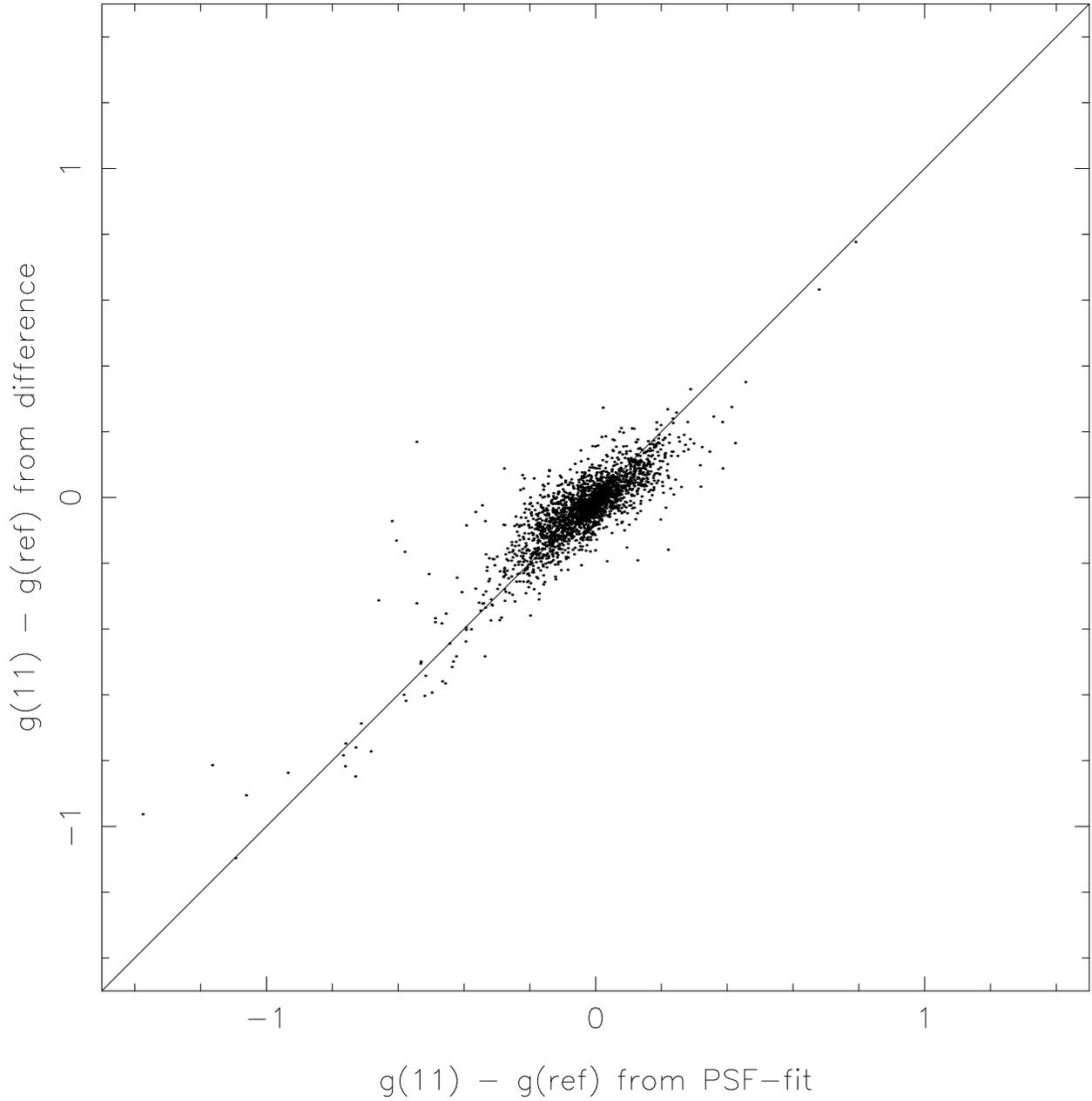}
\caption{A comparison of magnitude differences between epoch g11 and the reference image for stars on chip 1.  Only stars with signal-to-noise of at least 10 in both reductions are plotted.  The x axis shows $g$ magnitudes from the PSF-fitting photometry, while the y axis shows $g$ magnitudes calculated by our image subtraction code.  The diagonal line is $y=x$, not a fit to the data.}
\label{fig_diff}
\end{figure}

The difference images were then run through the HSTphot PSF-fitting routine, giving brightness differences (on the brightness scale of the epoch image rather than the reference image).  Scaling this difference to brightness scale of the reference image and adding the star's brightness in the reference image, the routine provides magnitudes directly comparable with magnitudes from the reference image.  Because this calculation is somewhat indirect (and thus conceivably error-prone), we show a comparison of epoch 11 magnitudes as calculated from image subtraction photometry and from standard PSF-fitting photometry in Figure \ref{fig_mags}.  We also show magnitude differences between epoch g11 and the reference image in Figure \ref{fig_diff}.  In both cases we find excellent agreement between the magnitudes measured directly using PSF-fitting photometry and those measured indirectly using image subtraction photometry.
\placefigure{fig_mags}
\placefigure{fig_diff}

In our variable star search described in the next section, there were two major advantages for using the image subtraction magnitudes.  First, as noted above, the photometry was significantly more accurate due to the better subtraction of extended objects and faint stars.  This allowed cleaner photometry of the RR Lyraes, which were more susceptible to these effects because of their faintness.  Not only does this make for generally better measurements of variability, but it made for fewer detections that had to be rejected because of fitting problems, thus giving more complete light curves.

A second significant difference is that slightly extended objects are treated more correctly.  Standard PSF-fitting photometry in HSTphot will treat an object as a star unless there is reason to believe otherwise.  This means that objects that are slightly extended will be photometered using a stellar PSF.  However, when the seeing varies (as it did during our run), these objects will be less affected by seeing variations, causing them to appear brighter in the poor-seeing images and thus producing an apparent light curve that tracks seeing changes.  In contrast, image subtraction will match the seeing in the image, and thus such an object will be subtracted out correctly.  While such ``seeing variables'' are easily weeded out due to their clear signature, using image subtraction gives a much cleaner initial sample.

\subsection{Variable Star Identification}

Our variable star identification procedure was very similar to that used in Leo A \citep{dol02}, so we will merely summarize here.  We began with a series of automatic selection steps, of which a star had to pass all to be flagged as a candidate variable.  A star had to pass a variety of steps to make our list of variables.  Our $g$ photometry list contained a total of 61008 objects, of which 26552 had reliable photometry (classified as a star and fit well in the PSF-fitting process) in at least 10 epochs.  From that list of stars, 665 were flagged as variable based on significant variability in the 16 epochs.

We then ran a periodicity test on these 665 stars to test for light curve coherence.  The statistic used was a variation on the $\theta$ statistic of \citet{laf65}, is defined by
\begin{equation}
\theta = \frac{\sum_i ( m_i - m_{i+1} ) ^2 }{\sum_i ( m_i - \mean{m} ) ^2 },
\end{equation}
where $m_i$ is the $i^{th}$ point on the light curve, $m_{i+1}$ is the next point, and $\mean{m}$ is the uncertainty-weighted mean magnitude.  (Here and below, point number $N+1$ is defined to be point number $1$.)  The denominator is, of course, equals the variance in $m$ times the number of observations.  If the period used to create the light curve is the correct period, then the quantity $(m_i - m_{i+1})^2$ will be much smaller than the variance; if the period is incorrect then $m_i$ and $m_{i+1}$ are merely two values drawn from a random sample and thus $(m_i - m_{i+1})^2$ should equal twice the variance.

The assumption made by this statistic is that the uncertainties of the magnitudes are constant (or at least close to constant).  However, in the case of our data, the drastic seeing variations, the factor of two variation in effective exposure times, and the signal-to-noise variations caused by variability itself cause the $\theta$ statistic to work rather poorly.  Rather than using the square of the magnitude differences, we prefer to have a statistic that is based on the significance of those differences.  The most similar to the theta statistic would be
\begin{equation}
\theta ' = \frac{\sum_i ( m_i - m_{i+1} ) ^2 / ( \sigma_i^2 + \sigma_{i+1}^2 ) }{\sum_i ( m_i - \mean{m} ) ^2 / \sigma_i^2 },
\end{equation}
where $\sigma_i$ is the uncertainty in $m_i$ and $\mean{m}$ here is the uncertainty-weighted mean magnitude (weight = $1/\sigma^2$).  The denominator, of course, is simply $\chi^2$ for a fit to constant magnitude.  The advantage of this statistic is that a bad point with high uncertainty is given the low weight that it deserves.  However, it introduces a failure mode in which highly uncertain points will preferentially be used as bridges between high and low points in the light curve -- bright point, uncertain point, faint point (or vice versa).

To combat this problem, we chose to examine coherence between more distant points on the light curve, rather than just between adjacent points.  The statistic we have derived to accomplish this is
\begin{equation}
\theta ' = \frac{ \sum_i \sum_{j \not= i} w_{ij} ( m_i - m_j ) ^2 / ( \sigma_i^2 + \sigma_j^2 ) }{ \chi_{\nu}^2 \sum_i \sum_{j \not= i} w_{ij} },
\end{equation}
\begin{equation}
w_{ij} = \left\{ \begin{array}{ll}
         1 & \mbox{if $j = i+1$};\\
         \cos( \phi_j - \phi_i ) & \mbox{if $j \not= i+1$ and $0 \leq \phi_j - \phi_i < 0.25$};\\
         0 & \mbox{otherwise},\end{array} \right.
\end{equation}
where $\phi_i$ is the phase of point $i$ and varies between 0 and $2 \pi$ and $\chi_{\nu}^2$ is the reduced $\chi^2$.  The quantity $\theta '$ will be one if $m_i$ and $m_j$ are uncorrelated since the variance of $m_i - m_j$ equals $\chi_{\nu}^2 ( \sigma_i^2 + \sigma_j^2 )$.  If the period is correct, the variance of $m_i - m_j$ should equal $\sigma_i^2 + \sigma_j^2 + A^2 ( 1 - \cos(\phi_j - \phi_i) )$, where $A$ is related to the amplitude of the variability.  (For sinusoidal variability, $A$ equals half the peak-to-peak variation; for other light curve shapes it will be different.)  For our data, we found that stars with $\theta ' \le 0.5$ have a high probability of being true variables, and thus used that cutoff.  Applying this criterion, we had 246 remaining candidate variables from the initial list of 665.

We note that there are other options for testing periodicity.  An obvious one would be a Fourier transform, but we found this to be more strongly affected by period aliasing and by bad points than our $\theta '$ statistic.

After paring the list to 246 candidate variables by automatic checks, we made a set of manual checks of the variability and periodicity.  Objects were examined on the individual images to verify that they were not extended and were indeed variable, and the light curves were verified visually.  This reduced the number of variable stars to 37.  Most of our false detections were cases in which a cosmic ray landed on the star in one or two epochs, or stars near the edge in which positions were not correctly mapped from the reference image to the epoch.  A summary of the properties of these 37 bona fide variable stars are given in Table \ref{tab_variables}, with complete photometry given in Table 3 and light curves in Figure \ref{fig_LC}.  A finding chart is shown in Figure \ref{fig_chart}.
\placetable{tab_variables}
\placetable{tab_phot}
\placefigure{fig_LC}
\placefigure{fig_chart}

\begin{figure}
\plotone{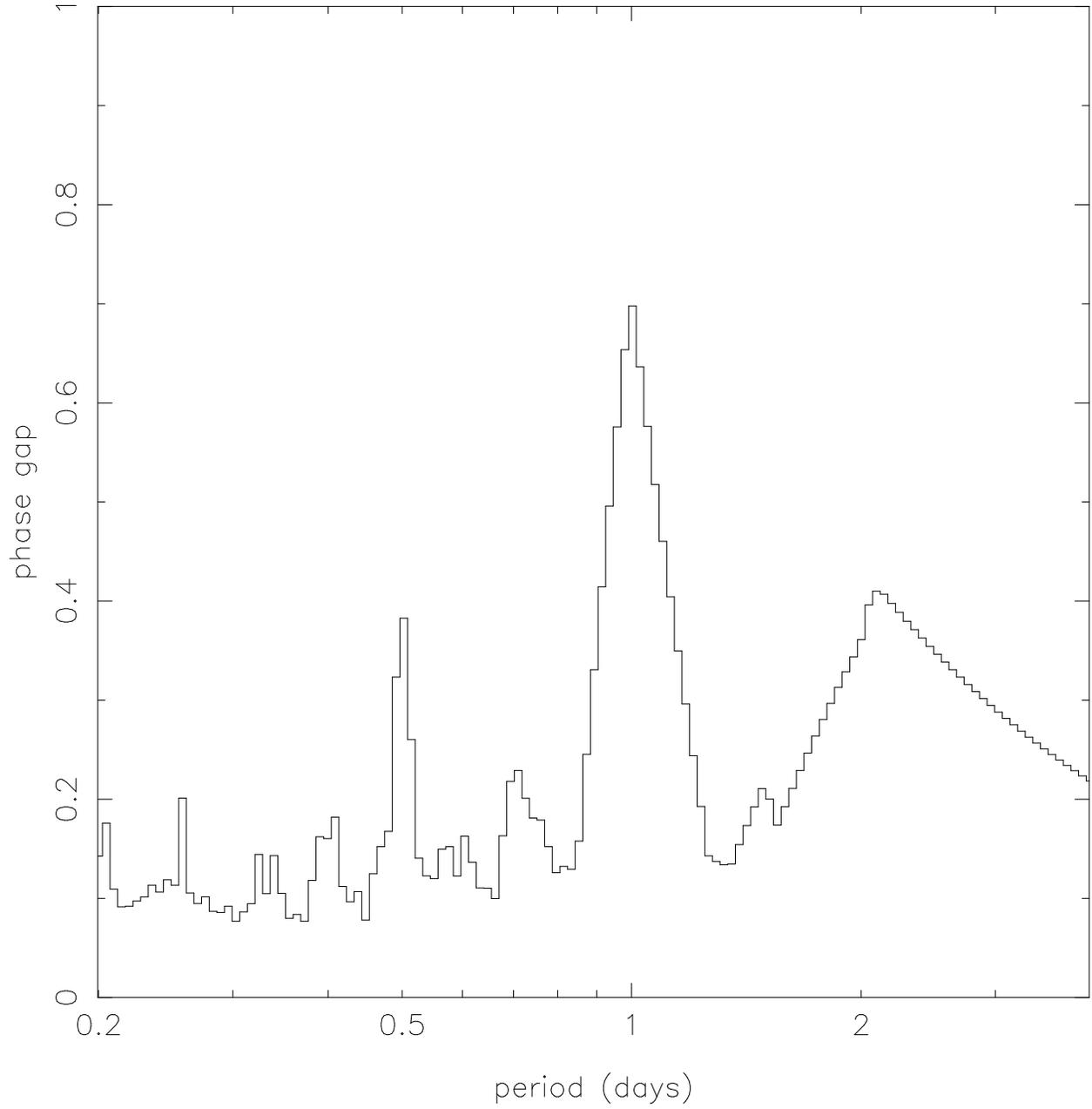}
\caption{Maximum gap in $g$ phase coverage, as a function of period.  Note that a variable star with a period of one day would have had under 30\% of its light curve covered by our observations.  A similar problem occurs for stars with periods near half a day.}
\label{fig_window}
\end{figure}

As is typical for work with ground-based data, stars with periods near one day had only a small fraction of their light curves covered by our observations.  We show our phase coverage as a function of period in Figure \ref{fig_window}.  This figure also indicates that we had poor phase coverage for stars with periods near half a day, again caused by the 8-hour window during which observations were made.  The lack of adequate phase coverage affects both discovery completeness and aliasing in determining periods for objects that are discovered.  Our period analysis found many RR Lyraes that could have equally likely had periods of one half or one third of a day; this degeneracy was unfortunate given that an RR Lyrae can have either period.  For these stars and others with ambiguous periods, we provide both options in Table \ref{tab_variables} and Figure \ref{fig_LC}.
\placefigure{fig_window}

\section{Census of RR Lyraes}

Among our list of 37 variable star candidates are 24 likely RR Lyraes.  Most likely the thirteen remaining variables are a mixture of anomalous Cepheids, population II Cepheids, and blended RR Lyraes.  However we have unambiguous periods for only six of the thirteen and $r$ magnitudes for only four of those six; thus definitive classification is impossible.

In order to estimate the actual number of RR Lyraes in the field, we have to estimate our incompleteness, which has three major sources.  First, as indicated in the previous paragraph, an RR Lyrae with a period of between 0.48 and 0.52 days would only have $60-70\%$ of its light curve covered by our $g$ photometry and thus would be less likely to make our list of variables.  Since no RR Lyraes with periods in this range were recovered, we assume a completeness of close to zero for such stars.  If stars with periods between 0.48 and 0.52 days account for approximately 12\% of RR Lyraes as in the OGLE LMC database \citep{uda99}, this factor would create an incompleteness of $\sim 12\%$.

The second source of incompleteness is that a minority of our images were sufficiently deep to obtain adequate photometry of RR Lyraes near their minimum.  Our two epochs with excellent seeing (g02 and g03) were separated by 0.1 days, and our only other epoch with seeing better than 0.75 arcsec (g04) was obtained 0.05 days after g03.  The photometry in the two best epochs was complete to $V \sim 26.6$, approximately a half magnitude deeper than the average of the other epochs.  Dividing our list of 24 RR Lyraes into four groups based on phase at g03 (with equal phase range in each group), we find that nearly half (11) were at minimum during that epoch.  If we interpret this as a signal of incompleteness, we estimate that the incompleteness caused by seeing changes was approximately 45\%.

Estimating that another 50\% of the RR Lyraes were thrown out during our conservative selection process, we calculate a total completeness of $\sim 24\%$ ($0.88 \times 0.55 \times 0.5$).  This implies that there are roughly 100 RR Lyraes in our field (a density of 1.1 RR Lyraes per square arcmin), an estimate that we believe accurate to a factor of two (the uncertainty due to estimations made in the completeness correction).

For comparison, PvdB claimed the discovery of 30 RR Lyraes with a completeness fraction of 25\% in a 7.26 square arcminute field.  Extrapolating this number to our observed field, we should have expected to find $\sim 1520$ RR Lyraes.  Even if we were complete only at their level, we should have recovered $\sim 380$ RR Lyraes.  The fact that we were only able to find one fifteenth that number, despite better seeing and deeper photometry, is a source of concern.

A second estimate of the number of RR Lyraes can be taken from the recent deep ACS photometry by \citet{bro03} in a field 51 arcminutes from the nucleus of M31.  Although they show their CMD in greyscale and do not report magnitudes on a standard system, we estimate from their CMD that there are approximately ten stars in the RR Lyrae region of their CMD.  Scaling from the ACS field of view to ours and adjusting for the distance to the nucleus of M31, this translates to approximately 120 RR Lyraes in our field of view, which is in good agreement with the number we have estimated.

Since our field includes the entire PvdB field, we are able to examine more carefully the source of the discrepancy between our findings.  Table \ref{tab_pvdb} gives a list of the 32 variable stars they found.  Of these, fourteen were sufficiently extended or blended that we could not obtain stellar photometry in our template $g$ image.  A fifteenth fell on a bad column and could not be photometered.  Fourteen of the remaining seventeen failed to pass our variability criteria, all having $\chi^2 < 2$ for their individual-epoch $g$ magnitudes.  Of those fourteen, only number 3 has properties consistent with those of an Lyrae, with a blue color and moderate $\chi^2$ value of 1.8.  The others are redder and/or non-variable to the accuracy of our photometry.

Of the three stars in common between our lists, two appear too bright to be isolated RR Lyraes.  PvdB noted that star 7 (our V08) was too bright to be an RR Lyrae and felt it was likely a blend of an RR Lyrae with a fainter non-variable star.  We agree with this assessment, as its color appears too red to fall in the instability strip, making a blend with a fainter, redder star the most likely scenario.  A second common star, number 15 (our V21) is comparably bright (in their study and in ours), but its color places it in the instability strip and thus we believe it to be a population II Cepheid or an anomalous Cepheid.  The third common variable is star 17 (our V13), which appears to be a bona fide RR Lyrae.

Although we cannot directly compare our photometry because common bands were not used, we can do so indirectly.  The typical $B-g$ color of an RR Lyrae star should be approximately 0.35, including the foreground extinction to this field.  For the three matches between our photometry lists, we find a mean $B-g$ difference of $0.56 \pm 0.06$ magnitudes, implying a discrepancy of roughly 0.2 magnitudes (in the sense that our photometry is brighter).Including all well-photometered stars with similar $g-r$ colors as RR Lyraes, we find that a mean $B-g$ difference of $0.66 \pm 0.13$ magnitudes (the larger scatter most likely due to the larger color range).  The source of this uncertainty is unclear.  As described in section \ref{sec_obs}, two of us independently reduced and photometered the data using different packages and agreed to one percent.  We also find the MIMO detector to have no measurable nonlinearity (better than 0.02 magnitude over 5 magnitudes; Saha et al. in prep), eliminating a second potential source of error.  We note also that \citet{pri88} found that the measured $\mean{B-V}$ color was 0.2 magnitudes redder than expected, which could be explained by an error of $+0.2$ magnitudes in their $B$ magnitudes.  We thus will proceed under the assumption that our magnitudes are accurate.

Comparing the RR Lyrae population with the number of red giants with magnitudes in the range $23.15 < g < 24.15$, we calculate a ratio of 30 red giants per RR Lyrae (assuming 24\% completeness).  For comparison, an RR Lyrae-rich globular cluster such as M3 has a ratio of approximately 6 \citep{sah92}.  Adopting M3 as an upper limit on the RR Lyrae production rate and using a conservative completeness correction, we conclude that at least 10\% of M31 red giants are drawn from an ancient, RR Lyrae-producing population.

\section{RR Lyrae Metallicities}

Averaging all 24 RR Lyrae candidates, we find a mean magnitude of $\mean{g} = 25.15 \pm 0.03$ (rms scatter of $\pm 0.14$ magnitudes).  If the five marginal-quality RR Lyraes (quality of 2 in Table \ref{tab_variables}) are removed, these numbers are unchanged.  From the 13 RR Lyraes with $r$ magnitudes, we find a mean color of $\mean{g-r} = -0.10 \pm 0.05$ (rms scatter of $\pm 0.20$ magnitudes).

To correct for reddening, we used the extinction maps of \citet{sch98} to calculate $E(B-V) = 0.09 \pm 0.03$.  Applying the extinction curve of \citet{car89} with $R_V=3.1$ and $\lambda_g = 4930$\AA, we calculate $A_g = 1.143 A_V = 0.32 \pm 0.10$.  For an $r$ wavelength of $\lambda_g = 6550$\AA, we calculate $A_r = 0.820 A_V = 0.23 \pm 0.08$, giving $E(g-r) = 0.09 \pm 0.03$.  This value agrees well with the mean color of our RR Lyraes, which compared with a typical RR Lyrae color of $(g-r)_0 = -0.17$ (Saha et al. in prep) implies a reddening of $E(g-r) = 0.07 \pm 0.05$.  Correcting our mean RR Lyrae magnitude for extinction, we calculate $\mean{g_0} = 24.83 \pm 0.11$ ($\mean{V_0} = 24.81 \pm 0.11$).

Adopting a true M31 distance modulus of $(m-M) = 24.44 \pm 0.10$ \citep{mad91}, we calculate an absolute magnitude of the RR Lyraes of $\mean{M_V} = 0.37 \pm 0.15$.  This is significantly brighter than one would expect given a halo metallicity of $\logz = -0.8$ \citep{dur01} and the RR Lyrae absolute magnitude calibration of \citet{car00}:
\begin{equation}
M_V = (0.18 \pm 0.09) (\feh + 1.5) + (0.57 \pm 0.07) = 0.70 \pm 0.09.
\end{equation}

We find several possible explanations for the discrepancy: uncertainties in the RR Lyrae zero point, uncertainties in the adopted M31 distance, incompleteness in our photometry, or errors in the assumed metallicity or metallicity effects.  We find no evidence for a large error in the zero point, since our study of IC 1613 \citep{dol01} used a variety of distance indicators to measure an RR Lyrae absolute magnitude of $M_V = 0.61 \pm 0.08$ at $\logz = -1.3 \pm 0.2$, which is in excellent agreement with the \citet{car00} calibration at that metallicity.

It appears equally implausible that the adopted distance is in error.  \citet{mou86} report a distance modulus of $(m-M)_0 = 24.4 \pm 0.2$ from their red giant branch tip, while \citet{sta98} use the red clump to measure $(m-M)_0 = 24.47 \pm 0.06$.

A completeness bias is plausible, since the seeing changes during our run made us insensitive to faint RR Lyraes near maximum in our best-seeing epochs.  To test this hypothesis, we selected the eleven RR Lyraes whose phases maximized the odds of detection (at minimum in our best-seeing epochs) and thus minimize this possible bias.  Those RR Lyraes had a mean magnitude of $\mean{g} = 25.16 \pm 0.04$, which is not significantly fainter than the overall average.  We note that this test is imperfect, as the RR Lyraes tested are themselves ones that passed any selection biases; nevertheless we would have expected to see a difference of much more than 0.01 magnitudes between had our sample been significantly biased against faint RR Lyraes.  A comparison with the deep ACS CMD of \citet{bro03} also argues against large photometry or selection biases.  At the expected color of RR Lyraes, their horizontal branch falls at $m_{F606W} \approx 25.30$, which corresponds to $V \approx 25.15$.

This leaves the final option, metallicity, as the most reasonable scenario.  While \citet{dur01} and others find the M31 halo to be metal-rich on average, they found the metallicity distribution best-fit with a bimodal distribution: roughly 60\% of the stars in a narrow distribution centered at $\logz = -0.5$ and the remaining stars in a very broad distribution ($\sigma = 0.45$ dex) centered at $\logz = -1.2$.  If the rate of RR Lyrae production is a strong function of metallicity as suggested by \citet{tam76}, one would expect RR Lyraes to be preferentially drawn from the metal-poor part of this distribution.  Given our relatively large uncertainty of $\pm 0.15$ magnitudes and the large range of metallicity dependencies of $M_V(RR)$ in the literature (e.g. Sandage \& Cacciari 1990 calculated a slope of 0.37 mags per dex, compared with 0.15 mags per dex from Carney, Storm, \& Jones 1992), our observations are consistent with the RR Lyraes being drawn from the metal-poor portion of the \citet{dur01} distribution.

We also note that variations in the RR Lyrae absolute magnitude have only been thoroughly examined for Galactic globular clusters, which fall within a very limited range of the possible age-metallicity space.  In particular, there are few (if any) young, metal-poor globular clusters.  Likewise, use of equation 5 makes implicit assumptions regarding the helium \citep{cia77} and alpha element \citep{van00} abundances as a function of iron abundance, which are not necessarily the same in the M31 halo as they are in Galactic globulars.  Therefore our conclusions regarding the implied metallicity of M31 halo RR Lyraes are only strictly correct if such stars have ages and abundance ratios comparable to those of Galactic globular clusters.

\section{Summary}

We have presented the results of a search for RR Lyraes in the M31 halo using observations taken over four nights on the WIYN 3.5m telescope.  Our $9.6 \times 9.6$ arcminute field was selected to include the smaller field used by a previous study by \citet{pri87}, which was 40 arcminutes southeast of M31's nucleus along the minor axis.  Applying a combination of PSF-fitting and image difference photometry techniques, we were able to detect 37 variable stars, including 24 RR Lyraes.

We estimate an completeness fraction of 24\%, which implies that roughly 100 RR Lyraes are present in this field -- slightly more than 1 RR Lyrae per square arcminute.  This number is much less than that calculated by \citet{pri87} ($\sim 17$ per square arc minute), and is consistent with the frequency of RR Lyraes expected from deep HST color-magnitude diagrams \citep{bro03}.

The mean magnitude of the RR Lyraes is $\mean{g} = 25.15 \pm 0.03$, which is consistent with the accepted M31 distance modulus of $(m-M) = 24.44$ only if RR Lyraes have a significantly lower metallicity than the mean value of $\logz = -0.8$ \citep{dur01}.  We interpret this as an indication that RR Lyraes are predominantly created by an old, low-metallicity population within the M31 halo (i.e. very similar to the population that appears to dominate the Milky Way halo) that accounts for at least 10\% of the red giants.

\acknowledgments
We would like to thank the NOAO TAC for the allocation of telescope time and the KPNO staff for their assistance.  We wish to thank our referee, Chris Pritchet, for helpful comments.  EO was partially supported by the NSF under grant AST 0098518.

\clearpage
\begin{figure}
\caption{Combined $g$ image of our field, centered near $0^h 45^m 32^s$, $+40^{\circ} 50' 30''$ (J2000).  North is to the right and East is up.  Each chip contains $2048 \times 4096$ pixels, with a scale of 0.141 arcsec per pixel.  The field area is 9.6 arcminutes on a side.}
\label{fig_image}
\end{figure}

\begin{figure}
\caption{$(g-r)$, $g$ CMD of the M31 halo from the deep photometry of the reference images.  Variable stars with qualities of 3 or 4 are plotted; RR Lyraes as asterisks and other variables as open circles.}
\label{fig_cmd}
\end{figure}

\begin{figure}
\caption{A comparison of the photometry for the first epoch, as measured by DoPHOT and HSTphot.}
\label{fig_photcomp}
\end{figure}

\begin{figure}
\caption{Illustration of our image subtraction technique.  The top two panels show part of the epoch g11 image centered on V12 (left) and the same part of the reference image.  The bottom left panel shows the photometry residuals from PSF-fitting photometry, while the bottom-right panel shows the difference image.  All images are shown on the same scale.  What is significant is that extended objects (background galaxies) are not subtracted very well by standard PSF-fitting techniques, leaving residuals that are seen in the bottom-left panel.  However, these objects are subtracted by the image subtraction technique, allowing a clean image on which the variable star can be photometered.}
\label{fig_res}
\end{figure}

\clearpage
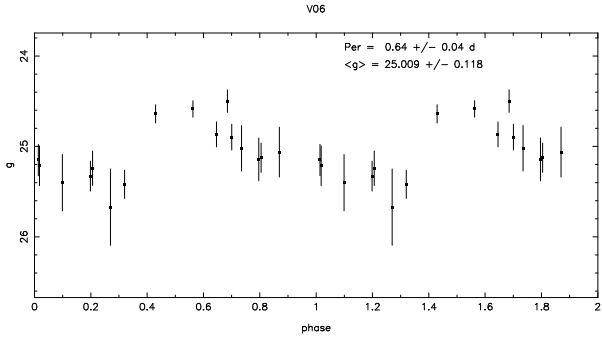
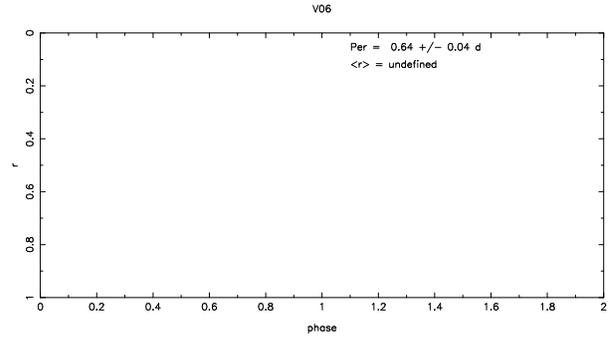
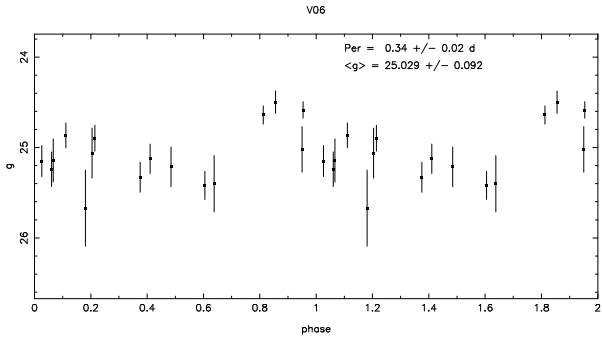
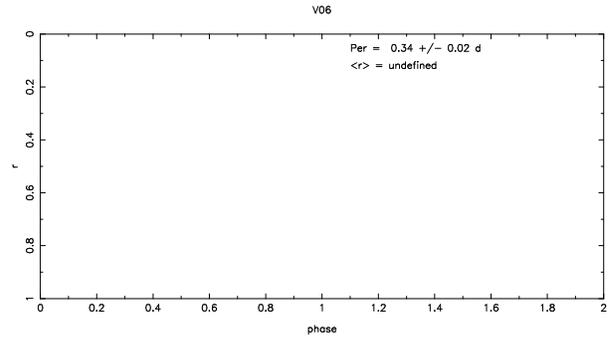
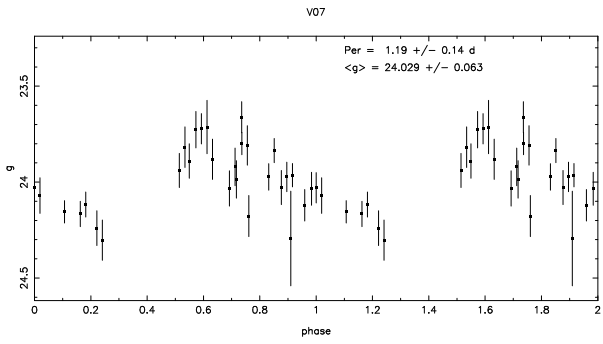
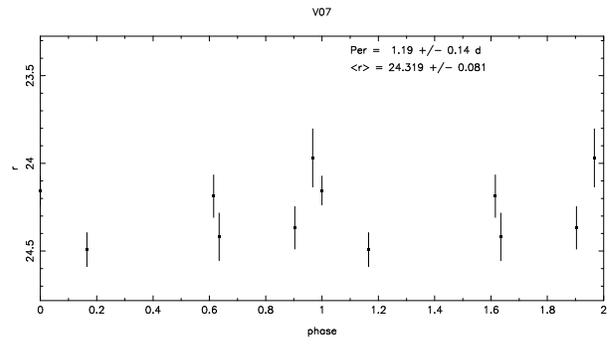
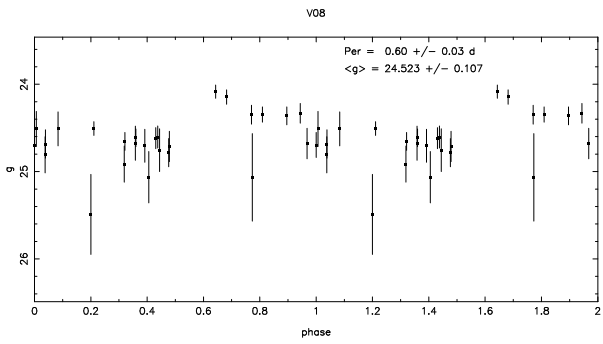
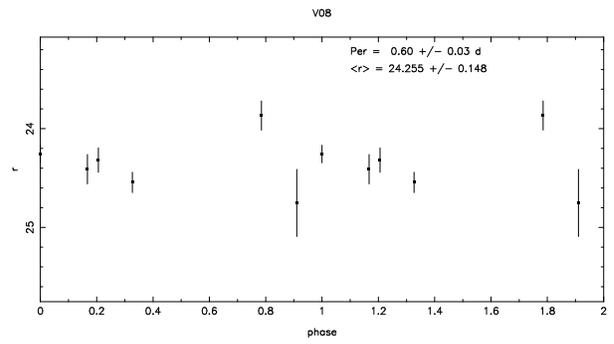
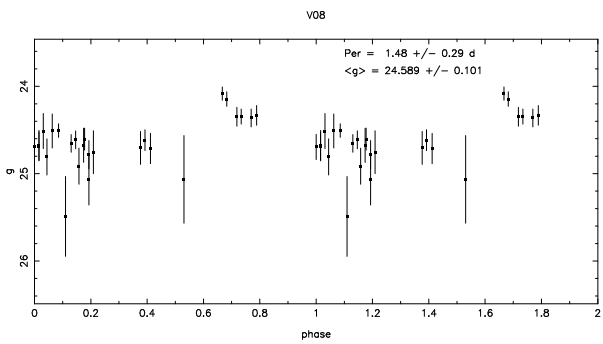
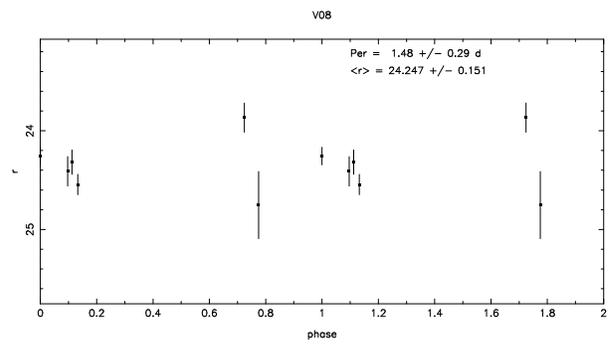
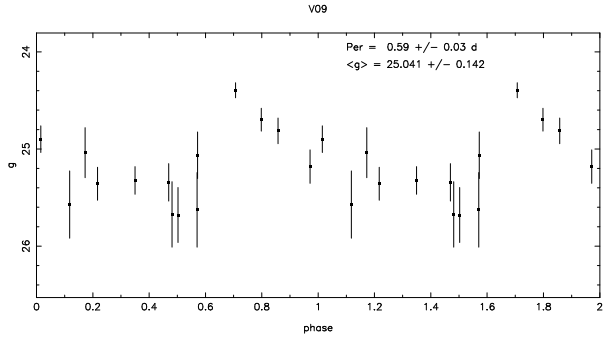
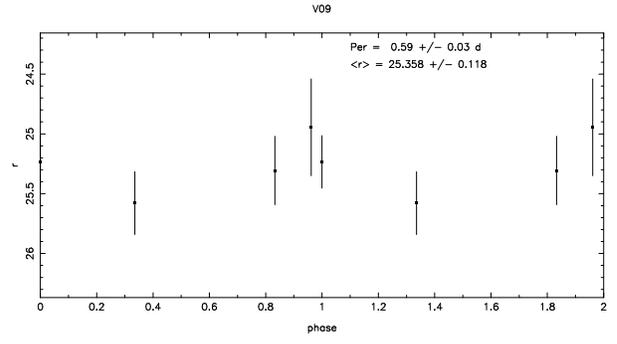
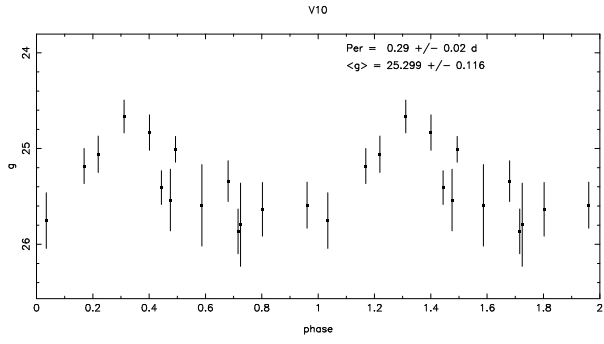
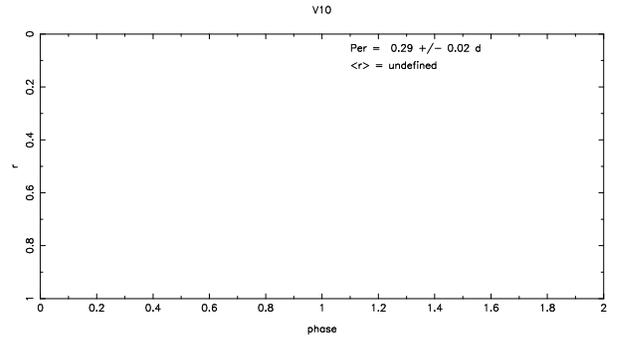
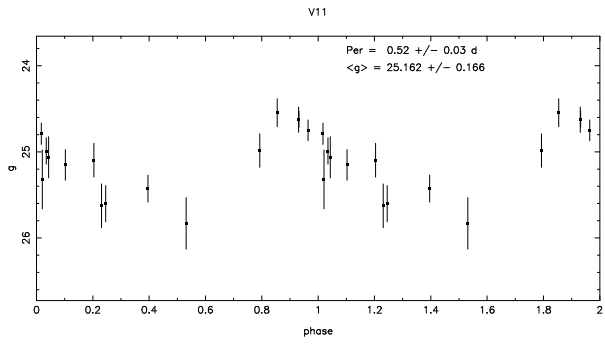
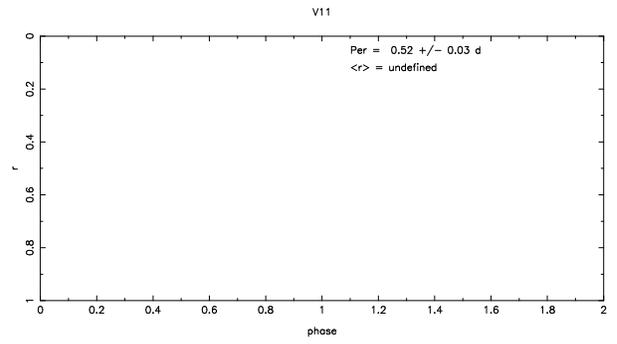
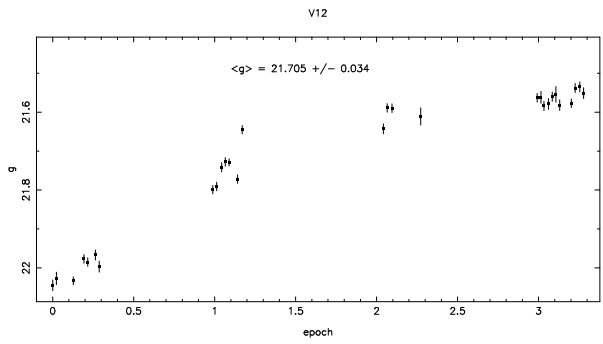
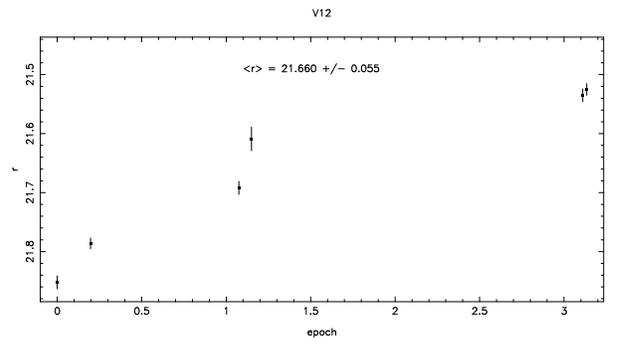
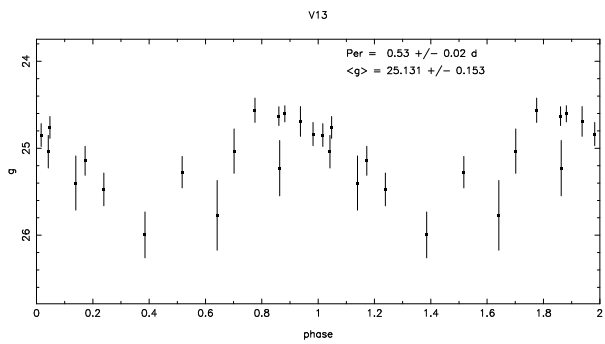
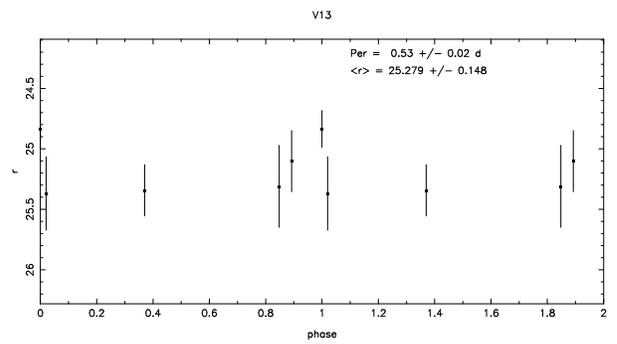
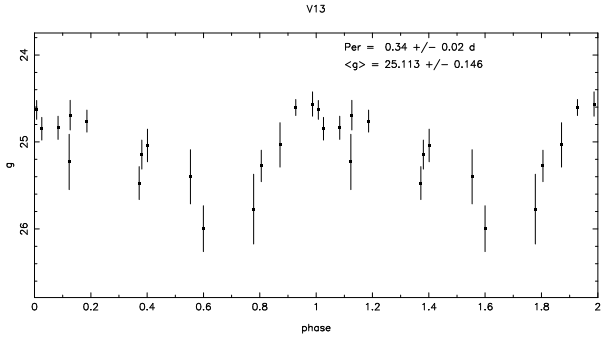
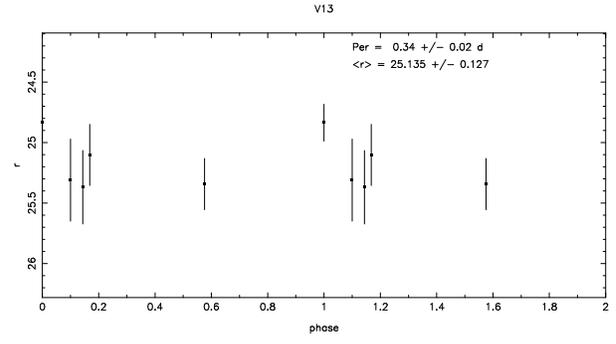
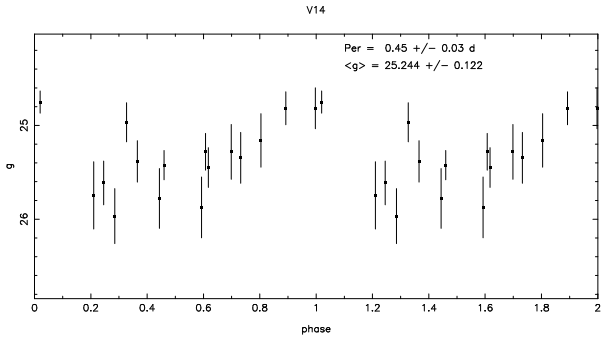
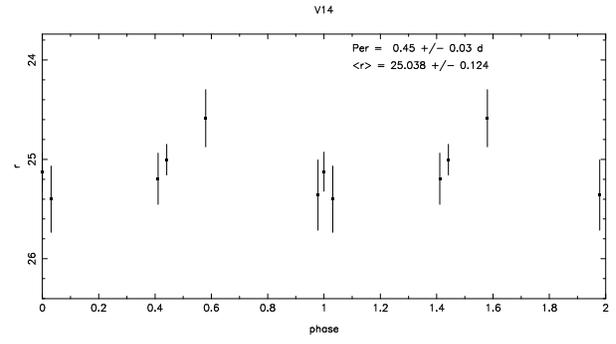
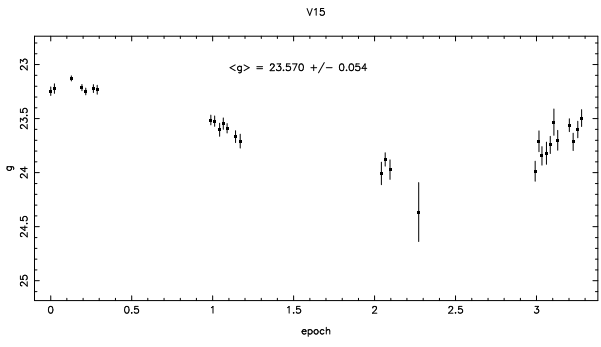
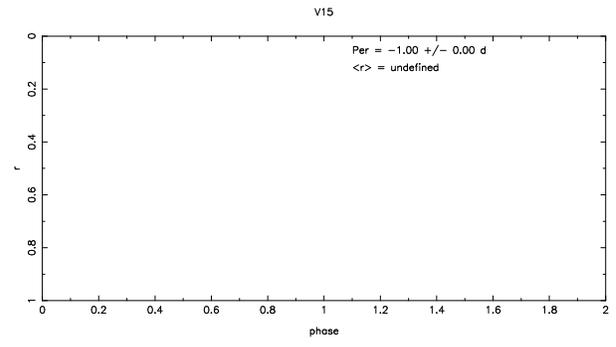
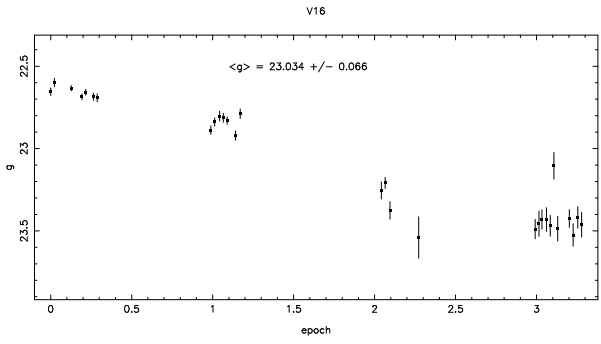
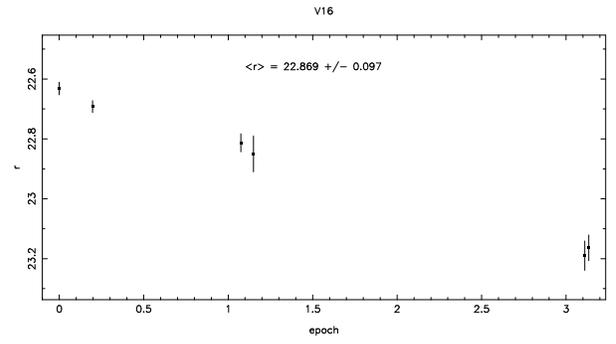
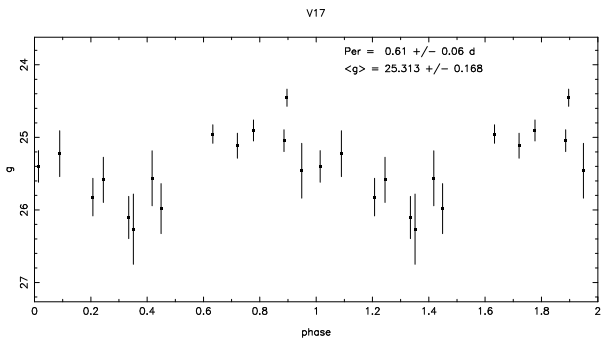
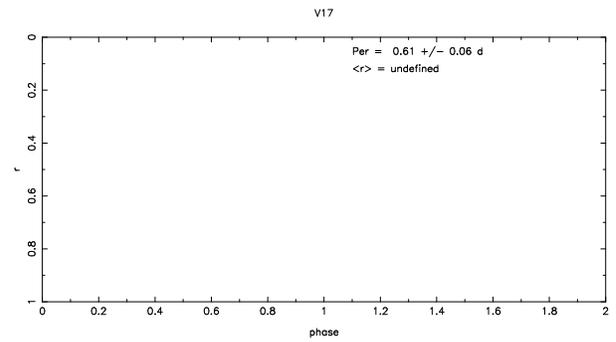
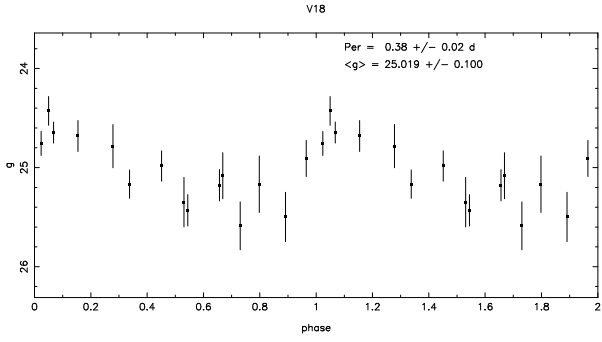
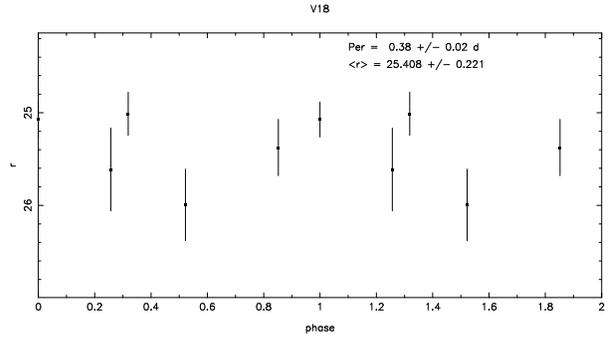
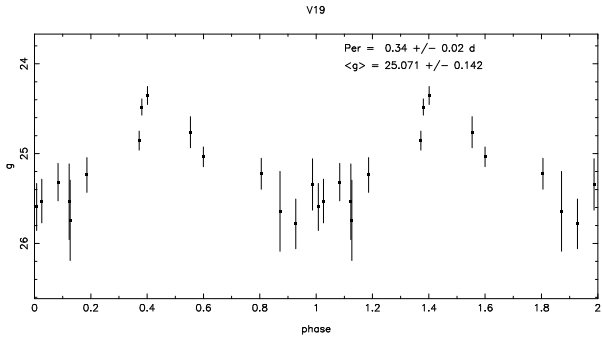
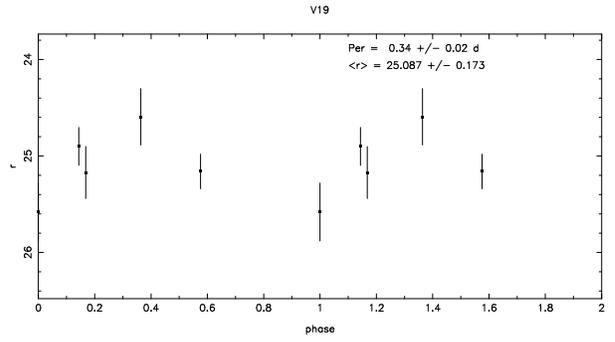
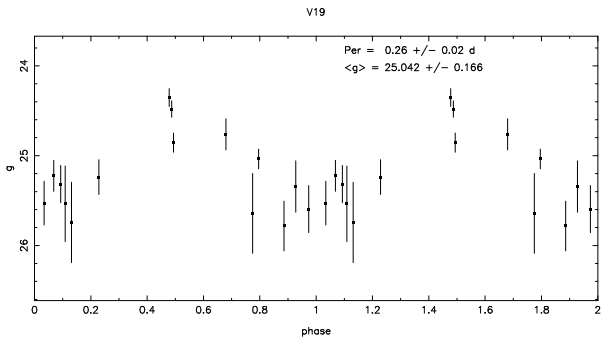
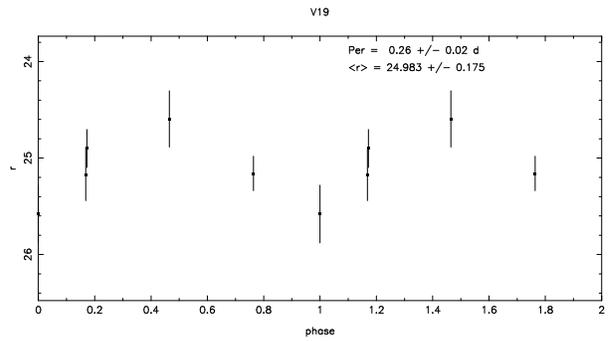
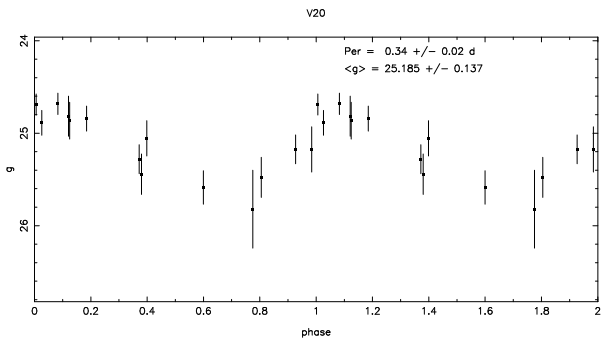
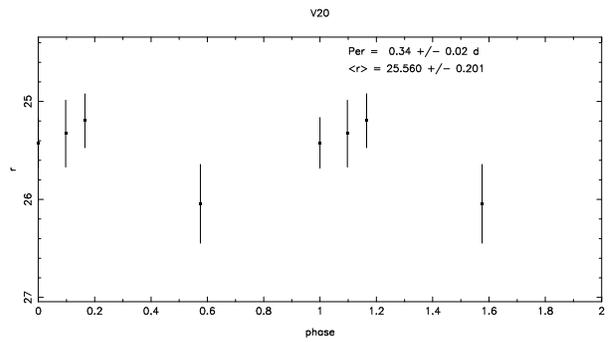
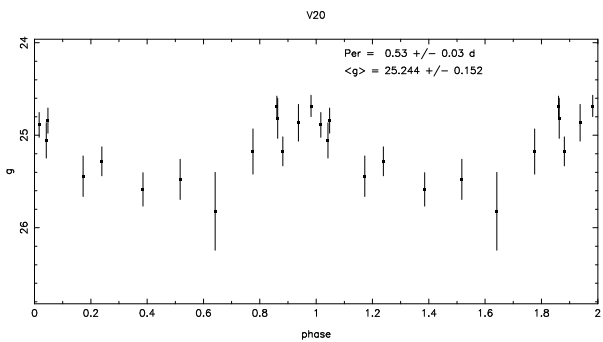
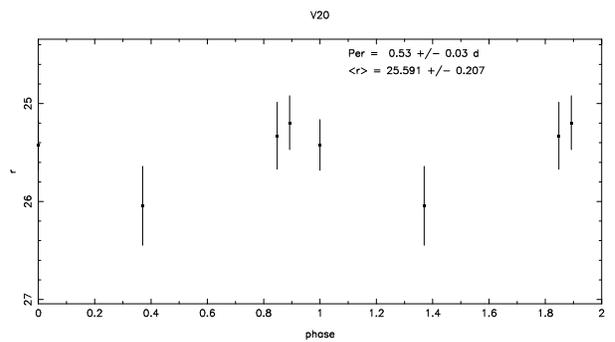
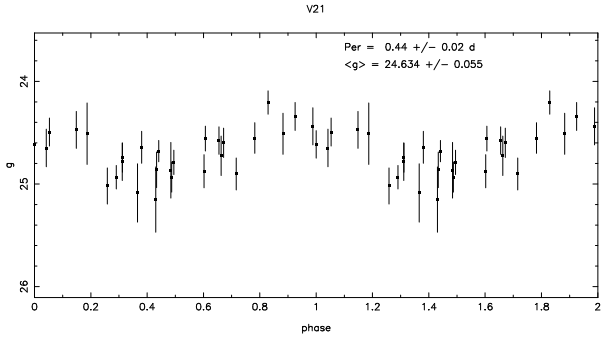
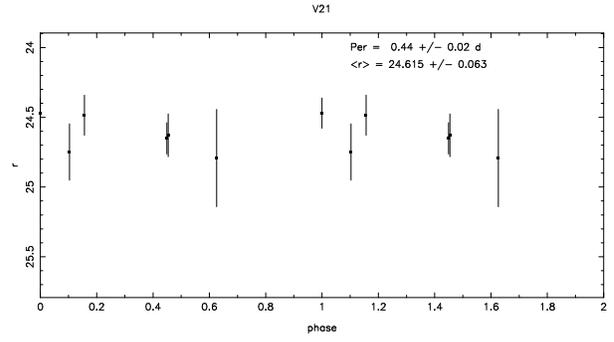
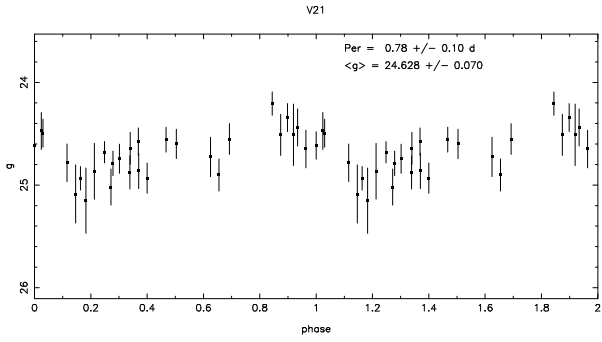
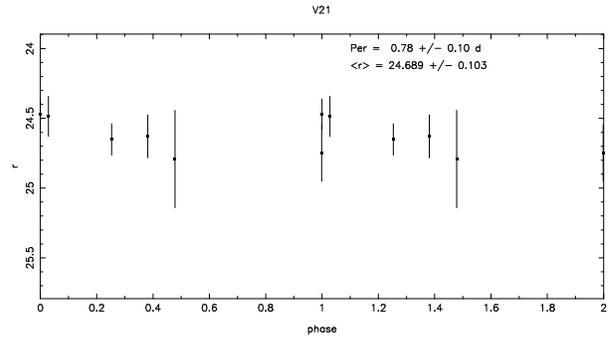
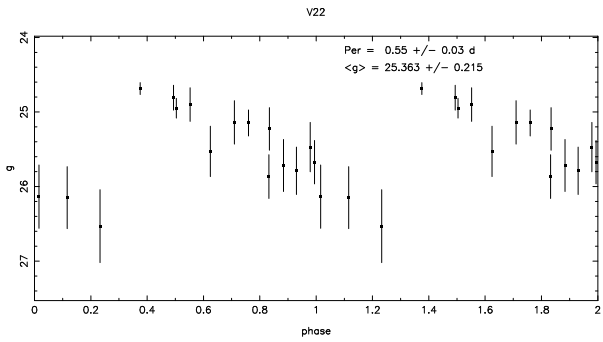
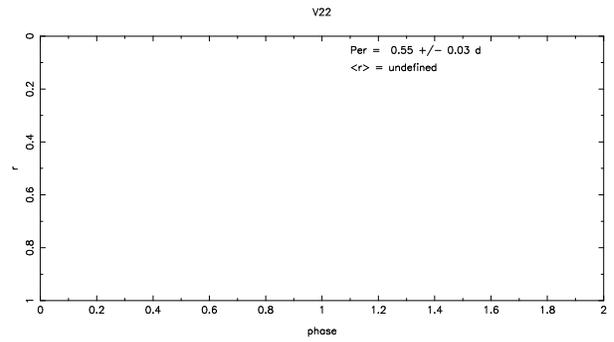
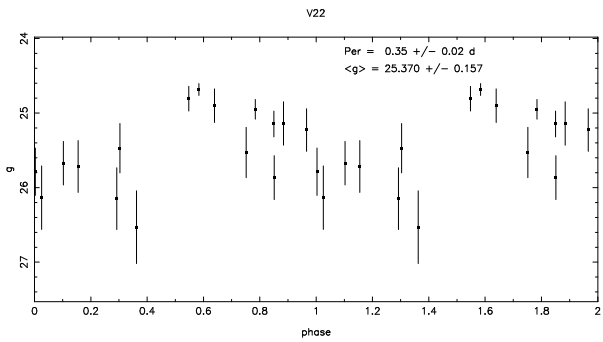
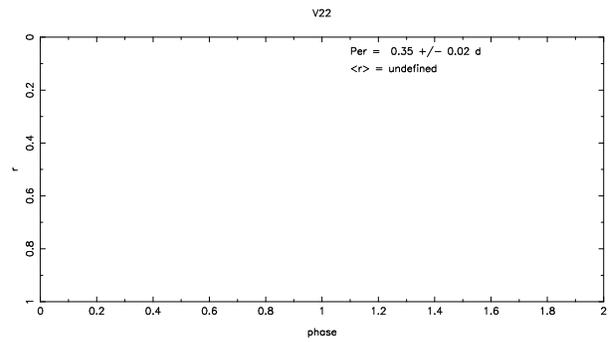
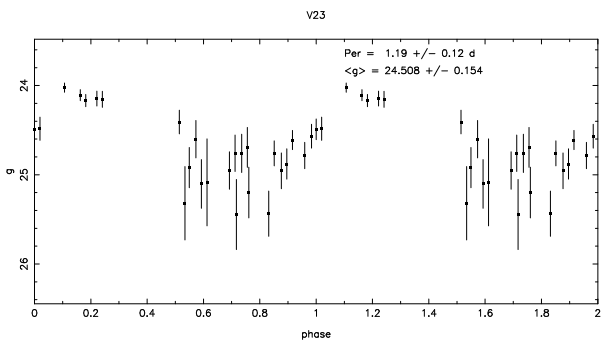
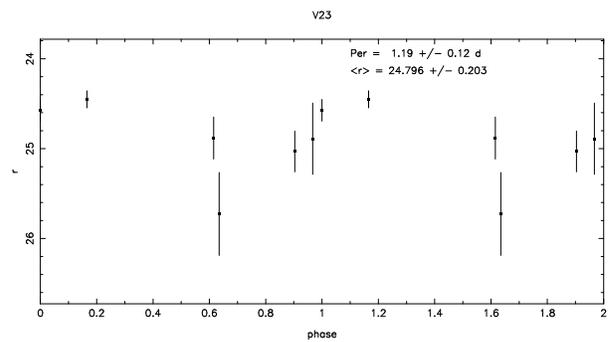
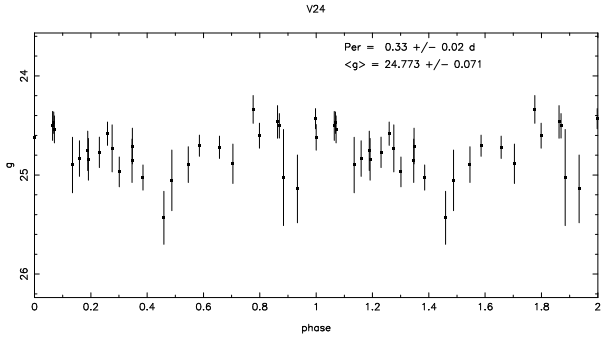
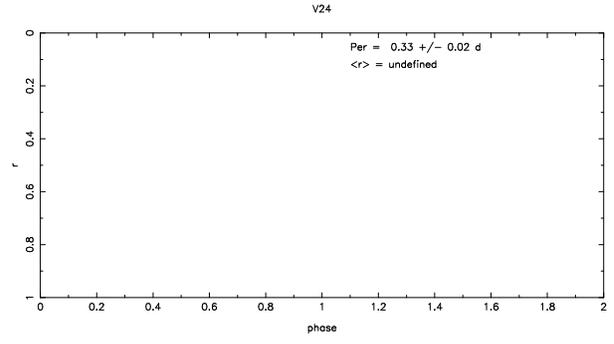

\clearpage

\def\fnum@figure{{\rmfamily Fig.\space\thefigure.---}}%
\clearpage
\begin{figure}
\plotone{dolphin.fig07a.ps}
\caption{Light curves of variable stars.  Magnitudes of non-periodic variables or variables whose periods are longer than our 3.3 day baseline are plotted vs. epoch (HJD-2452230.6) rather than vs. phase.  For all variables with mean magnitudes brighter than $\mean{g} = 24.9$, we plot $g$ magnitudes from the single image photometry; for those fainter than that limit the $g$ magnitudes are from our deeper coadded pairs of images.}
\label{fig_LC}
\end{figure}

\clearpage
\begin{figure}
\plotone{dolphin.fig07b.ps}
\addtocounter{figure}{-1}
\caption{Continued}
\end{figure}

\clearpage
\begin{figure}
\plotone{dolphin.fig07c.ps}
\addtocounter{figure}{-1}
\caption{Continued}
\end{figure}

\clearpage
\begin{figure}
\plotone{dolphin.fig07d.ps}
\addtocounter{figure}{-1}
\caption{Continued}
\end{figure}

\clearpage
\begin{figure}
\plotone{dolphin.fig07e.ps}
\addtocounter{figure}{-1}
\caption{Continued}
\end{figure}

\clearpage
\begin{figure}
\plotone{dolphin.fig07f.ps}
\addtocounter{figure}{-1}
\caption{Continued}
\end{figure}

\clearpage
\begin{figure}
\plotone{dolphin.fig07g.ps}
\addtocounter{figure}{-1}
\caption{Continued}
\end{figure}

\clearpage
\begin{figure}
\plotone{dolphin.fig07h.ps}
\addtocounter{figure}{-1}
\caption{Continued}
\end{figure}

\clearpage
\begin{figure}
\plotone{dolphin.fig07i.ps}
\addtocounter{figure}{-1}
\caption{Continued}
\end{figure}

\clearpage
\begin{figure}
\plotone{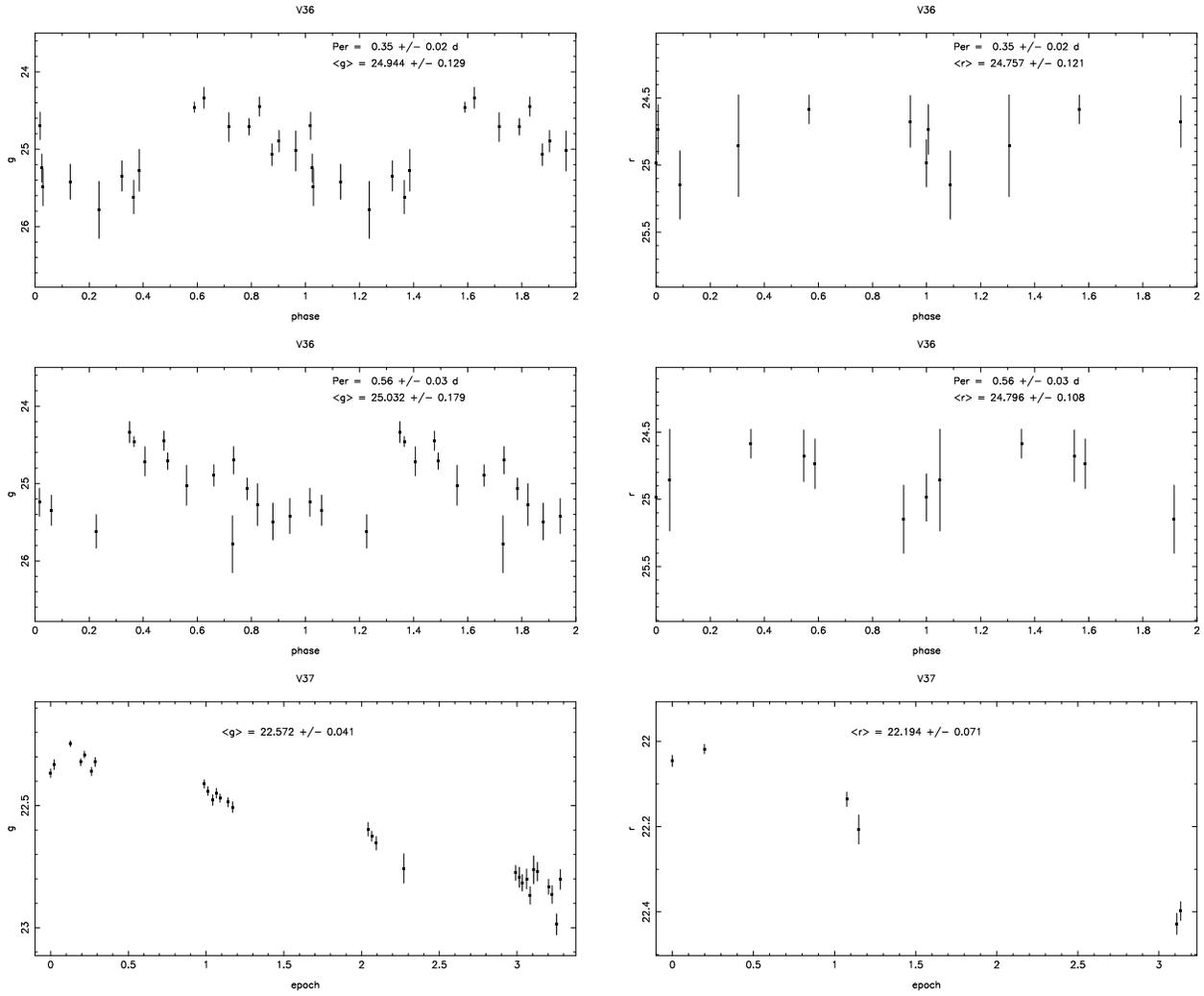}
\addtocounter{figure}{-1}
\caption{Continued}
\end{figure}

\clearpage
\begin{figure}
\caption{Positions of 37 variables in our M31 halo field.}
\label{fig_chart}
\end{figure}

\clearpage

\clearpage
\begin{deluxetable}{lcccccc}
\tablecaption{Observation Log. \label{tab_obslog}}
\tablewidth{0pt}
\tablehead{
\colhead{Image} & Epoch & \colhead{HJD\tablenotemark{a}} & \colhead{Exposure} & \colhead{Filter} & \colhead{Airmass} & \colhead{Seeing} \\
\colhead{ID} & \colhead{} & \colhead{} & \colhead{Time (s)} & \colhead{} & \colhead{} & \colhead{(arcsec)\tablenotemark{b}}
}
\startdata
n1050 & g01 & 230.60021 & 1800 & $g$ & 1.15 & 0.84 \\
n1051 & g01 & 230.61947 & 1100 & $g$ & 1.09 & 0.78 \\
n1064 & g02 & 230.72735 & 1800 & $g$ & 1.02 & 0.52 \\
n1071 & g03 & 230.79348 & 1800 & $g$ & 1.14 & 0.59 \\
n1072 & g03 & 230.81702 & 1800 & $g$ & 1.22 & 0.58 \\
n1074 & g04 & 230.86383 & 1800 & $g$ & 1.48 & 0.70 \\
n1075 & g04 & 230.88714 & 1800 & $g$ & 1.69 & 0.74 \\
n2028 & g05 & 231.58951 & 1800 & $g$ & 1.18 & 0.80 \\
n2029 & g05 & 231.61281 & 1800 & $g$ & 1.11 & 0.77 \\
n2031 & g06 & 231.64287 & 1800 & $g$ & 1.05 & 0.93 \\
n2032 & g06 & 231.66617 & 1800 & $g$ & 1.02 & 0.85 \\
n2033 & g07 & 231.68948 & 1800 & $g$ & 1.01 & 0.77 \\
n2036 & g08 & 231.74181 & 1800 & $g$ & 1.04 & 0.83 \\
n2038 & g08 & 231.77033 & 1800 & $g$ & 1.09 & 0.79 \\
n3030 & g09 & 232.64245 & 1800 & $g$ & 1.05 & 0.85 \\
n3031 & g09 & 232.66576 & 1800 & $g$ & 1.02 & 0.85 \\
n3033 & g10 & 232.69478 & 1800 & $g$ & 1.01 & 0.94 \\
n4031 & g11 & 233.59133 & 1800 & $g$ & 1.16 & 0.88 \\
n4032 & g11 & 233.61198 & 1340 & $g$ & 1.09 & 1.03 \\
n4033 & g12 & 233.63327 & 1800 & $g$ & 1.06 & 0.99 \\
n4035 & g13 & 233.66107 & 1800 & $g$ & 1.02 & 1.18 \\
n4036 & g13 & 233.68437 & 1800 & $g$ & 1.01 & 1.09 \\
n4037 & g14 & 233.70768 & 1800 & $g$ & 1.02 & 1.05 \\
n4038 & g14 & 233.73107 & 1800 & $g$ & 1.03 & 0.92 \\
n4041 & g15 & 233.80249 & 1800 & $g$ & 1.19 & 0.97 \\
n4042 & g15 & 233.82634 & 1800 & $g$ & 1.30 & 1.05 \\
n4044 & g16 & 233.85477 & 1800 & $g$ & 1.47 & 1.09 \\
n4045 & g16 & 233.87811 & 1800 & $g$ & 1.68 & 1.12 \\
\tableline
n1052 & r01 & 230.64364 & 1800 & $r$ & 1.05 & 0.68 \\
n1073 & r02 & 230.84043 & 1800 & $r$ & 1.33 & 0.58 \\
n2035 & r03 & 231.71832 & 1800 & $r$ & 1.02 & 0.80 \\
n2039 & r04 & 231.79372 & 1800 & $r$ & 1.15 & 0.71 \\
n4039 & r05 & 233.75470 & 1800 & $r$ & 1.07 & 1.01 \\
n4040 & r06 & 233.77801 & 1800 & $r$ & 1.12 & 0.96 \\
\enddata
\tablenotetext{a}{HJD$-$2452000.}
\tablenotetext{b}{Seeing values taken from our PSF-fitting photometry.}
\end{deluxetable}

\clearpage
\begin{deluxetable}{lrrccccc}
\tablecaption{Variable Stars. \label{tab_variables}}
\tablewidth{0pt}
\tablehead{
\colhead{ID\tablenotemark{a}} &
\colhead{X} &
\colhead{Y} &
\colhead{g} &
\colhead{r} &
\colhead{P (d)\tablenotemark{b}} &
\colhead{Q\tablenotemark{c}} &
\colhead{Class\tablenotemark{d}}}
\startdata
V01 &  388.02 & 2314.30 & $25.15 \pm 0.14$ & $25.43\pm0.15$ & $0.59\pm0.02$ & 4 & RR    \\
V02 &  406.11 &  287.99 & $25.40 \pm 0.15$ & $25.47\pm0.33$ & $0.53\pm0.02$ & 4 & RR    \\
V03 &  590.52 & 2535.80 & $24.92 \pm 0.09$ & $24.97\pm0.12$ & $0.62\pm0.02$ & 4 & RR    \\
V04 &  683.70 & 2418.53 & $24.99 \pm 0.10$ &     . . .      & $0.58\pm0.03$ & 4 & RR    \\
V05 &  875.57 & 3542.01 & $25.24 \pm 0.13$ &     . . .      & $0.36\pm0.02$ & 3 & RR    \\
V06 & 1085.89 & 2667.73 & $25.01 \pm 0.12$ &     . . .      & $0.64\pm0.04$ & 4 & RR    \\
V06 & 1085.89 & 2667.73 & $25.03 \pm 0.09$ &     . . .      & $0.34\pm0.02$ & 4 & RR    \\
V07 & 1094.93 & 1806.49 & $24.03 \pm 0.06$ & $24.32\pm0.08$ & $1.19\pm0.14$ & 3 & --    \\
V08 & 1147.99 & 1934.58 & $24.52 \pm 0.11$ & $24.25\pm0.15$ & $0.60\pm0.03$ & 4 & --    \\
V08 & 1147.99 & 1934.58 & $24.59 \pm 0.10$ & $24.25\pm0.15$ & $1.48\pm0.29$ & 4 & --    \\
V09 & 1163.53 & 1113.15 & $25.04 \pm 0.14$ & $25.36\pm0.12$ & $0.59\pm0.03$ & 4 & RR    \\
V10 & 1173.81 & 3395.22 & $25.30 \pm 0.12$ &     . . .      & $0.29\pm0.02$ & 2 & RR    \\
V11 & 1231.75 &  565.27 & $25.16 \pm 0.17$ &     . . .      & $0.52\pm0.03$ & 4 & RR    \\
V12 & 1368.03 & 1267.04 & $21.70 \pm 0.03$ & $21.66\pm0.06$ &     . . .     & 4 & --    \\
V13 & 1539.16 & 1699.02 & $25.13 \pm 0.15$ & $25.28\pm0.15$ & $0.53\pm0.02$ & 3 & RR    \\
V13 & 1539.16 & 1699.02 & $25.11 \pm 0.15$ & $25.14\pm0.13$ & $0.34\pm0.02$ & 3 & RR    \\
V14 & 1597.99 & 1029.53 & $25.24 \pm 0.12$ & $25.04\pm0.12$ & $0.45\pm0.03$ & 3 & RR    \\
V15 & 1614.25 & 3177.27 & $23.57 \pm 0.05$ &     . . .      &     . . .     & 4 & --    \\
V16 & 1655.14 & 2805.79 & $23.03 \pm 0.07$ & $22.87\pm0.10$ &     . . .     & 4 & --    \\
V17 & 1786.01 &  940.75 & $25.31 \pm 0.17$ &     . . .      & $0.61\pm0.06$ & 2 & RR    \\
V18 & 1853.16 & 3355.98 & $25.02 \pm 0.10$ & $25.41\pm0.22$ & $0.38\pm0.02$ & 3 & RR    \\
V19 & 1882.30 & 1753.31 & $25.07 \pm 0.14$ & $25.09\pm0.17$ & $0.34\pm0.02$ & 2 & RR    \\
V19 & 1882.30 & 1753.31 & $25.04 \pm 0.17$ & $24.98\pm0.17$ & $0.26\pm0.02$ & 2 & RR    \\
V20 & 1928.36 &  923.53 & $25.18 \pm 0.14$ & $25.56\pm0.20$ & $0.34\pm0.02$ & 4 & RR    \\
V20 & 1928.36 &  923.53 & $25.24 \pm 0.15$ & $25.59\pm0.21$ & $0.53\pm0.03$ & 4 & RR    \\
V21 & 1974.12 & 1209.61 & $24.63 \pm 0.06$ & $24.61\pm0.06$ & $0.44\pm0.02$ & 4 & --    \\
V21 & 1974.12 & 1209.61 & $24.63 \pm 0.07$ & $24.69\pm0.10$ & $0.78\pm0.10$ & 3 & --    \\
V22 & 2153.33 &  558.56 & $25.36 \pm 0.21$ &     . . .      & $0.55\pm0.03$ & 3 & RR    \\
V22 & 2153.33 &  558.56 & $25.37 \pm 0.16$ &     . . .      & $0.35\pm0.02$ & 3 & RR    \\
V23 & 2324.98 & 3187.69 & $24.51 \pm 0.15$ & $24.80\pm0.20$ & $1.19\pm0.12$ & 3 & --    \\
V24 & 2365.18 & 2420.98 & $24.77 \pm 0.07$ &     . . .      & $0.33\pm0.02$ & 2 & --    \\
V25 & 2784.98 & 2978.03 & $25.12 \pm 0.13$ &     . . .      & $0.33\pm0.02$ & 2 & RR    \\
V26 & 2789.02 & 1804.45 & $25.02 \pm 0.14$ & $25.13\pm0.13$ & $0.23\pm0.02$ & 4 & RR    \\
V26 & 2789.02 & 1804.45 & $25.14 \pm 0.16$ & $25.04\pm0.10$ & $0.46\pm0.02$ & 4 & RR    \\
V27 & 3027.28 & 1557.52 & $25.27 \pm 0.12$ & $25.34\pm0.18$ & $0.34\pm0.02$ & 3 & RR    \\
V27 & 3027.28 & 1557.52 & $25.26 \pm 0.14$ & $25.31\pm0.18$ & $0.51\pm0.02$ & 3 & RR    \\
V28 & 3063.67 & 3289.28 & $25.24 \pm 0.14$ &     . . .      & $0.56\pm0.04$ & 4 & RR    \\
V29 & 3086.78 & 1080.11 & $24.95 \pm 0.13$ &     . . .      & $0.24\pm0.02$ & 2 & RR    \\
V29 & 3086.78 & 1080.11 & $24.90 \pm 0.11$ &     . . .      & $0.32\pm0.02$ & 2 & RR    \\
V30 & 3116.35 & 1713.88 & $25.24 \pm 0.15$ &     . . .      & $0.35\pm0.02$ & 3 & RR    \\
V31 & 3182.21 &  597.45 & $23.30 \pm 0.06$ & $23.31\pm0.11$ &     . . .     & 4 & --    \\
V32 & 3254.82 & 1416.38 & $24.64 \pm 0.10$ &     . . .      & $0.61\pm0.04$ & 4 & --    \\
V33 & 3276.99 &  932.46 & $24.49 \pm 0.07$ & $23.64\pm0.05$ & $0.89\pm0.08$ & 2 & --    \\
V34 & 3314.54 & 1319.22 & $24.70 \pm 0.07$ & $24.80\pm0.08$ & $0.33\pm0.02$ & 3 & --    \\
V35 & 3322.44 & 1872.86 & $25.14 \pm 0.13$ & $25.09\pm0.06$ & $0.62\pm0.04$ & 4 & RR    \\
V36 & 3335.96 & 1768.64 & $24.94 \pm 0.13$ & $24.76\pm0.12$ & $0.35\pm0.02$ & 4 & RR    \\
V36 & 3335.96 & 1768.64 & $25.03 \pm 0.18$ & $24.80\pm0.11$ & $0.56\pm0.03$ & 4 & RR    \\
V37 & 3950.40 & 1700.69 & $22.57 \pm 0.04$ & $22.19\pm0.07$ &     . . .     & 4 & --    \\
\enddata
\tablenotetext{a}{Stars listed multiple times have periods with multiple potential aliases.}
\tablenotetext{b}{Variables listed without periods are either non-periodic or have periods longer than our 3.3 day baseline.}
\tablenotetext{c}{Variable star quality is based on the cleanness of the light curves and quality of the photometry, where 4 is highest quality and 0 is lowest quality.}
\tablenotetext{d}{RR: RR Lyrae, --: unclassified}
\end{deluxetable}

\clearpage
\begin{table}
\begin{center}
\addtocounter{table}{1}
Table \thetable.  Photometry of Variable Stars. \label{tab_phot}
\begin{tabular}{cccccc}
\\
\tableline
\tableline
        & V01            & V02            & V03            & V04            & V05             \\
\tableline
HJD$^a$  & g              & g              & g              & g              & g              \\
230.6523 & $25.55\pm0.25$ & $25.91\pm0.34$ &     . . .      & $25.25\pm0.19$ & $24.61\pm0.11$ \\
230.7706 & $25.07\pm0.13$ & $25.84\pm0.26$ & $25.02\pm0.12$ & $25.25\pm0.15$ & $25.24\pm0.15$ \\
230.8485 & $25.18\pm0.13$ & $25.90\pm0.24$ & $24.54\pm0.07$ & $25.30\pm0.14$ & $26.01\pm0.27$ \\
230.9188 & $24.62\pm0.10$ & $25.93\pm0.33$ & $24.49\pm0.09$ & $24.80\pm0.12$ & $25.67\pm0.26$ \\
231.6445 & $25.21\pm0.16$ & $24.84\pm0.11$ & $24.87\pm0.12$ & $24.71\pm0.10$ &     . . .      \\
231.6978 &     . . .      & $25.43\pm0.25$ & $25.19\pm0.18$ & $24.83\pm0.13$ & $25.13\pm0.18$ \\
231.7328 & $25.41\pm0.23$ & $25.35\pm0.22$ & $24.83\pm0.13$ & $25.05\pm0.16$ & $24.71\pm0.12$ \\
231.7994 & $26.01\pm0.36$ & $25.83\pm0.30$ & $25.27\pm0.18$ & $25.02\pm0.15$ & $25.01\pm0.15$ \\
232.6975 & $24.39\pm0.09$ & $25.11\pm0.17$ & $24.60\pm0.11$ & $24.54\pm0.10$ & $25.70\pm0.29$ \\
232.7382 & $24.46\pm0.14$ & $25.04\pm0.39$ & $24.60\pm0.16$ & $24.61\pm0.16$ & $25.84\pm0.50$ \\
233.6447 & $25.84\pm0.43$ & $25.80\pm0.41$ & $25.30\pm0.26$ & $25.42\pm0.29$ & $25.46\pm0.30$ \\
233.6767 & $25.63\pm0.50$ & $24.79\pm0.20$ & $25.63\pm0.44$ & $25.27\pm0.32$ &     . . .      \\
233.7161 & $25.89\pm0.48$ & $24.73\pm0.16$ & $25.17\pm0.24$ & $25.90\pm0.47$ & $25.30\pm0.28$ \\
233.7628 & $25.48\pm0.41$ & $25.09\pm0.28$ & $25.21\pm0.31$ & $25.01\pm0.26$ & $25.32\pm0.36$ \\
233.8578 & $24.39\pm0.11$ & $25.64\pm0.32$ & $25.19\pm0.21$ & $24.88\pm0.16$ & $25.70\pm0.34$ \\
233.9099 & $24.78\pm0.18$ & $25.20\pm0.26$ & $24.67\pm0.16$ & $24.66\pm0.16$ & $24.72\pm0.17$ \\
\tableline
HJD$^a$  & r              & r              & r              & r              & r              \\
230.6435 & $25.47\pm0.27$ &     . . .      & $25.34\pm0.24$ &     . . .      &     . . .      \\
230.8403 & $25.04\pm0.16$ & $26.16\pm0.45$ & $24.84\pm0.13$ &     . . .      &     . . .      \\
231.7182 & $25.68\pm0.40$ & $25.24\pm0.27$ & $24.81\pm0.18$ &     . . .      &     . . .      \\
231.7936 &     . . .      &     . . .      &     . . .      &     . . .      &     . . .      \\
233.7547 & $25.76\pm0.52$ &     . . .      &     . . .      &     . . .      &     . . .      \\
233.7780 & $25.21\pm0.28$ & $25.06\pm0.24$ & $25.40\pm0.34$ &     . . .      &     . . .      \\
\end{tabular}
\end{center}
$^a$ HJD$-$2452000
\end{table}
\clearpage
\begin{table}
\begin{center}
Table \thetable---Continued \\
\begin{tabular}{cccccc}
\\
\tableline
\tableline
        & V06            & V07            & V08            & V09            & V10             \\
\tableline
HJD$^a$  & g              & g              & g              & g              & g              \\
230.6523 & $25.15\pm0.17$ & $24.05\pm0.06$ & $24.62\pm0.11$ & $24.90\pm0.14$ &     . . .      \\
230.7706 & $25.33\pm0.17$ & $24.16\pm0.06$ & $24.51\pm0.08$ & $25.36\pm0.17$ & $25.41\pm0.18$ \\
230.8485 & $25.42\pm0.16$ & $24.15\pm0.05$ & $24.64\pm0.08$ & $25.32\pm0.14$ & $25.86\pm0.23$ \\
230.9188 & $24.64\pm0.10$ & $24.27\pm0.07$ & $24.71\pm0.11$ & $25.34\pm0.19$ & $25.59\pm0.24$ \\
231.6445 & $24.59\pm0.09$ & $23.93\pm0.05$ & $24.11\pm0.06$ & $24.39\pm0.08$ & $25.01\pm0.14$ \\
231.6978 & $24.87\pm0.14$ & $23.99\pm0.06$ & $24.23\pm0.10$ & $24.70\pm0.12$ & $25.34\pm0.21$ \\
231.7328 & $24.90\pm0.14$ & $23.96\pm0.06$ & $24.35\pm0.09$ & $24.81\pm0.13$ & $25.64\pm0.28$ \\
231.7994 & $25.12\pm0.16$ & $24.08\pm0.06$ & $24.33\pm0.08$ & $25.18\pm0.17$ & $25.75\pm0.29$ \\
232.6975 & $25.24\pm0.19$ & $23.86\pm0.05$ & $24.58\pm0.10$ & $25.68\pm0.28$ & $25.18\pm0.18$ \\
232.7382 & $25.67\pm0.42$ & $24.18\pm0.11$ & $24.71\pm0.17$ & $25.06\pm0.24$ & $24.66\pm0.17$ \\
233.6447 & $24.50\pm0.12$ & $23.90\pm0.07$ & $24.55\pm0.13$ & $25.57\pm0.35$ & $25.54\pm0.32$ \\
233.6767 & $25.02\pm0.25$ & $23.89\pm0.09$ & $24.81\pm0.21$ & $25.04\pm0.26$ & $25.59\pm0.43$ \\
233.7161 & $25.14\pm0.24$ & $23.70\pm0.06$ & $24.75\pm0.17$ &     . . .      & $25.80\pm0.43$ \\
233.7628 & $25.06\pm0.28$ & $23.82\pm0.09$ & $25.25\pm0.32$ &     . . .      &     . . .      \\
233.8578 & $25.21\pm0.22$ & $23.96\pm0.07$ & $24.76\pm0.14$ & $25.67\pm0.34$ & $25.06\pm0.19$ \\
233.9099 & $25.40\pm0.31$ & $23.77\pm0.07$ & $24.88\pm0.19$ & $25.63\pm0.38$ & $24.83\pm0.18$ \\
\tableline
HJD$^a$  & r              & r              & r              & r              & r              \\
230.6435 &     . . .      & $24.16\pm0.08$ & $24.26\pm0.09$ & $25.23\pm0.22$ &     . . .      \\
230.8403 &     . . .      & $24.49\pm0.10$ & $24.54\pm0.10$ & $25.58\pm0.26$ &     . . .      \\
231.7182 &     . . .      & $24.37\pm0.12$ & $23.87\pm0.15$ & $25.31\pm0.29$ &     . . .      \\
231.7936 &     . . .      & $23.97\pm0.17$ & $24.75\pm0.34$ & $24.95\pm0.41$ &     . . .      \\
233.7547 &     . . .      & $24.19\pm0.12$ & $24.41\pm0.15$ &     . . .      &     . . .      \\
233.7780 &     . . .      & $24.42\pm0.14$ & $24.32\pm0.12$ &     . . .      &     . . .      \\
\end{tabular}
\end{center}
\end{table}
\clearpage
\begin{table}
\begin{center}
Table \thetable---Continued \\
\begin{tabular}{cccccc}
\\
\tableline
\tableline
        & V11            & V12            & V13            & V14            & V15             \\
\tableline
HJD$^a$  & g              & g              & g              & g              & g              \\
230.6523 & $24.79\pm0.12$ & $22.04\pm0.01$ & $24.85\pm0.13$ & $24.75\pm0.12$ & $23.24\pm0.03$ \\
230.7706 & $25.60\pm0.21$ & $22.03\pm0.01$ & $25.47\pm0.19$ & $25.97\pm0.29$ & $23.13\pm0.02$ \\
230.8485 & $25.43\pm0.16$ & $21.98\pm0.01$ & $26.00\pm0.26$ & $25.42\pm0.16$ & $23.23\pm0.02$ \\
230.9188 & $25.83\pm0.30$ & $21.99\pm0.01$ & $25.27\pm0.18$ & $25.45\pm0.21$ & $23.21\pm0.03$ \\
231.6445 & $24.62\pm0.09$ & $21.80\pm0.01$ & $24.60\pm0.09$ & $25.61\pm0.23$ & $23.52\pm0.04$ \\
231.6978 & $24.99\pm0.15$ & $21.74\pm0.01$ & $24.84\pm0.13$ & $25.38\pm0.22$ & $23.56\pm0.04$ \\
231.7328 & $25.15\pm0.18$ & $21.73\pm0.01$ & $24.76\pm0.13$ & $25.78\pm0.32$ & $23.59\pm0.04$ \\
231.7994 & $25.63\pm0.26$ & $21.71\pm0.01$ & $25.14\pm0.17$ & $25.87\pm0.32$ & $23.64\pm0.04$ \\
232.6975 & $24.75\pm0.12$ & $21.61\pm0.01$ & $24.63\pm0.11$ & $25.28\pm0.20$ & $23.92\pm0.06$ \\
232.7382 & $25.06\pm0.24$ & $21.59\pm0.01$ & $24.69\pm0.17$ & $25.28\pm0.29$ & $23.97\pm0.09$ \\
233.6447 & $24.98\pm0.19$ & $21.57\pm0.01$ & $25.77\pm0.40$ & $25.35\pm0.27$ & $23.89\pm0.07$ \\
233.6767 & $24.55\pm0.16$ & $21.58\pm0.01$ & $25.03\pm0.26$ & $25.16\pm0.29$ & $23.84\pm0.09$ \\
233.7161 & $24.63\pm0.15$ & $21.57\pm0.01$ & $24.56\pm0.14$ & $24.82\pm0.18$ & $23.82\pm0.07$ \\
233.7628 & $25.32\pm0.34$ & $21.57\pm0.01$ & $25.23\pm0.32$ & $24.82\pm0.22$ & $23.61\pm0.07$ \\
233.8578 & $25.10\pm0.20$ & $21.56\pm0.01$ & $25.04\pm0.19$ & $25.75\pm0.36$ & $23.59\pm0.05$ \\
233.9099 &     . . .      & $21.54\pm0.01$ & $25.40\pm0.31$ & $24.97\pm0.21$ & $23.53\pm0.06$ \\
\tableline
HJD$^a$  & r              & r              & r              & r              & r              \\
230.6435 &     . . .      & $21.85\pm0.01$ & $24.84\pm0.15$ & $25.12\pm0.20$ &     . . .      \\
230.8403 &     . . .      & $21.79\pm0.01$ & $25.34\pm0.21$ & $25.00\pm0.15$ &     . . .      \\
231.7182 &     . . .      & $21.69\pm0.01$ & $25.37\pm0.30$ & $25.20\pm0.26$ &     . . .      \\
231.7936 &     . . .      & $21.61\pm0.02$ &     . . .      & $24.59\pm0.29$ &     . . .      \\
233.7547 &     . . .      & $21.54\pm0.01$ & $25.31\pm0.34$ & $25.36\pm0.35$ &     . . .      \\
233.7780 &     . . .      & $21.52\pm0.01$ & $25.10\pm0.25$ & $25.40\pm0.33$ &     . . .      \\
\end{tabular}
\end{center}
\end{table}
\clearpage
\begin{table}
\begin{center}
Table \thetable---Continued \\
\begin{tabular}{cccccc}
\\
\tableline
\tableline
        & V16            & V17            & V18            & V19            & V20             \\
\tableline
HJD$^a$  & g              & g              & g              & g              & g              \\
230.6523 & $22.63\pm0.02$ & $25.40\pm0.22$ & $24.76\pm0.12$ & $25.53\pm0.24$ & $24.89\pm0.14$ \\
230.7706 & $22.63\pm0.02$ & $25.82\pm0.26$ & $25.17\pm0.14$ & $24.86\pm0.11$ & $25.28\pm0.16$ \\
230.8485 & $22.67\pm0.01$ & $26.10\pm0.29$ & $25.43\pm0.16$ & $25.04\pm0.11$ & $25.59\pm0.18$ \\
230.9188 & $22.67\pm0.02$ & $25.98\pm0.34$ & $25.59\pm0.24$ & $25.22\pm0.17$ & $25.48\pm0.22$ \\
231.6445 & $22.85\pm0.02$ & $24.95\pm0.13$ & $25.18\pm0.16$ & $25.78\pm0.28$ & $25.17\pm0.16$ \\
231.6978 & $22.80\pm0.02$ & $25.11\pm0.17$ & $25.17\pm0.28$ & $25.32\pm0.21$ & $24.68\pm0.12$ \\
231.7328 & $22.83\pm0.02$ & $24.91\pm0.14$ & $25.50\pm0.25$ & $25.24\pm0.19$ & $24.84\pm0.14$ \\
231.7994 & $22.86\pm0.02$ & $25.05\pm0.15$ & $24.65\pm0.11$ & $24.48\pm0.09$ & $25.44\pm0.22$ \\
232.6975 & $23.23\pm0.03$ & $26.27\pm0.48$ & $24.98\pm0.15$ & $25.59\pm0.26$ & $24.69\pm0.11$ \\
232.7382 & $23.38\pm0.05$ & $25.57\pm0.38$ &     . . .      & $25.74\pm0.45$ & $24.86\pm0.20$ \\
233.6447 & $23.48\pm0.05$ & $24.45\pm0.12$ & $24.91\pm0.18$ &     . . .      & $25.82\pm0.42$ \\
233.6767 & $23.43\pm0.06$ & $25.46\pm0.38$ & $24.43\pm0.15$ & $25.64\pm0.45$ &     . . .      \\
233.7161 & $23.42\pm0.05$ &     . . .      & $24.68\pm0.16$ & $25.34\pm0.29$ & $25.18\pm0.24$ \\
233.7628 & $23.29\pm0.06$ & $25.23\pm0.32$ & $24.78\pm0.22$ & $25.54\pm0.42$ & $24.82\pm0.22$ \\
233.8578 & $23.46\pm0.04$ & $25.59\pm0.31$ & $25.35\pm0.25$ & $24.35\pm0.10$ & $25.05\pm0.19$ \\
233.9099 & $23.40\pm0.05$ &     . . .      & $25.08\pm0.23$ & $24.76\pm0.17$ &     . . .      \\
\tableline
HJD$^a$  & r              & r              & r              & r              & r              \\
230.6435 & $22.63\pm0.02$ &     . . .      & $25.07\pm0.19$ & $25.58\pm0.30$ & $25.42\pm0.26$ \\
230.8403 & $22.69\pm0.02$ &     . . .      & $26.00\pm0.39$ & $25.16\pm0.18$ & $26.04\pm0.40$ \\
231.7182 & $22.81\pm0.03$ &     . . .      & $25.38\pm0.30$ & $24.90\pm0.20$ &     . . .      \\
231.7936 & $22.85\pm0.06$ &     . . .      &     . . .      & $24.59\pm0.29$ &     . . .      \\
233.7547 & $23.19\pm0.05$ &     . . .      & $25.61\pm0.45$ &     . . .      & $25.33\pm0.34$ \\
233.7780 & $23.16\pm0.04$ &     . . .      & $25.01\pm0.23$ & $25.17\pm0.27$ & $25.20\pm0.28$ \\
\end{tabular}
\end{center}
\end{table}
\clearpage
\begin{table}
\begin{center}
Table \thetable---Continued \\
\begin{tabular}{cccccc}
\\
\tableline
\tableline
        & V21            & V22            & V23            & V24            & V25             \\
\tableline
HJD$^a$  & g              & g              & g              & g              & g              \\
230.6523 & $24.57\pm0.10$ & $26.14\pm0.42$ & $24.45\pm0.09$ & $24.56\pm0.10$ & $25.44\pm0.22$ \\
230.7706 & $24.93\pm0.12$ & $26.53\pm0.49$ & $24.02\pm0.05$ & $25.03\pm0.13$ & $24.88\pm0.11$ \\
230.8485 & $24.73\pm0.08$ & $24.68\pm0.08$ & $24.13\pm0.05$ & $24.71\pm0.08$ & $25.15\pm0.12$ \\
230.9188 & $24.66\pm0.10$ & $24.95\pm0.13$ & $24.17\pm0.07$ & $24.55\pm0.09$ & $25.97\pm0.34$ \\
231.6445 & $24.88\pm0.12$ & $25.87\pm0.29$ & $25.06\pm0.14$ & $24.43\pm0.08$ & $25.55\pm0.22$ \\
231.6978 & $24.71\pm0.12$ & $25.79\pm0.32$ & $24.99\pm0.15$ & $24.79\pm0.12$ & $24.72\pm0.12$ \\
231.7328 & $24.93\pm0.15$ & $25.67\pm0.29$ & $24.61\pm0.11$ & $24.97\pm0.15$ & $24.97\pm0.15$ \\
231.7994 & $24.64\pm0.10$ & $26.15\pm0.41$ & $24.61\pm0.10$ & $25.08\pm0.15$ & $25.13\pm0.16$ \\
232.6975 & $24.74\pm0.12$ & $25.15\pm0.17$ & $24.92\pm0.15$ & $24.56\pm0.10$ & $24.91\pm0.14$ \\
232.7382 & $24.55\pm0.15$ & $25.23\pm0.29$ & $25.20\pm0.28$ & $24.71\pm0.18$ & $24.76\pm0.19$ \\
233.6447 & $24.26\pm0.10$ & $24.81\pm0.17$ & $24.57\pm0.13$ & $24.63\pm0.14$ & $24.55\pm0.13$ \\
233.6767 & $24.34\pm0.14$ & $24.90\pm0.23$ & $24.92\pm0.23$ & $24.84\pm0.21$ & $24.74\pm0.19$ \\
233.7161 & $24.54\pm0.14$ & $25.53\pm0.34$ & $24.89\pm0.18$ & $24.84\pm0.18$ & $24.91\pm0.19$ \\
233.7628 & $24.48\pm0.16$ & $25.14\pm0.29$ & $25.41\pm0.38$ & $25.09\pm0.28$ & $25.12\pm0.29$ \\
233.8578 & $24.90\pm0.17$ & $25.72\pm0.35$ & $24.96\pm0.17$ & $24.60\pm0.12$ & $25.47\pm0.28$ \\
233.9099 & $24.95\pm0.21$ & $25.47\pm0.33$ & $24.63\pm0.15$ & $24.73\pm0.17$ & $25.53\pm0.34$ \\
\tableline
HJD$^a$  & r              & r              & r              & r              & r              \\
230.6435 & $24.47\pm0.11$ &     . . .      & $24.57\pm0.12$ &     . . .      &     . . .      \\
230.8403 & $24.65\pm0.11$ &     . . .      & $24.45\pm0.10$ &     . . .      &     . . .      \\
231.7182 & $24.63\pm0.15$ &     . . .      & $25.03\pm0.23$ &     . . .      &     . . .      \\
231.7936 & $24.79\pm0.35$ &     . . .      & $24.89\pm0.40$ &     . . .      &     . . .      \\
233.7547 & $24.75\pm0.20$ &     . . .      & $24.88\pm0.23$ &     . . .      &     . . .      \\
233.7780 & $24.48\pm0.14$ &     . . .      & $25.73\pm0.46$ &     . . .      &     . . .      \\
\end{tabular}
\end{center}
\end{table}
\clearpage
\begin{table}
\begin{center}
Table \thetable---Continued \\
\begin{tabular}{cccccc}
\\
\tableline
\tableline
        & V26            & V27            & V28            & V29            & V30             \\
\tableline
HJD$^a$  & g              & g              & g              & g              & g              \\
230.6523 & $25.47\pm0.23$ & $25.69\pm0.28$ & $26.25\pm0.46$ & $25.58\pm0.25$ & $24.56\pm0.10$ \\
230.7706 & $24.84\pm0.11$ & $25.66\pm0.23$ & $25.36\pm0.17$ & $25.26\pm0.16$ & $25.51\pm0.20$ \\
230.8485 & $25.12\pm0.12$ & $24.76\pm0.09$ & $25.15\pm0.12$ & $24.46\pm0.07$ & $25.46\pm0.16$ \\
230.9188 & $25.54\pm0.23$ & $25.02\pm0.14$ & $24.90\pm0.12$ & $25.02\pm0.14$ & $25.44\pm0.21$ \\
231.6445 & $24.63\pm0.09$ & $25.77\pm0.28$ & $25.35\pm0.18$ & $25.14\pm0.15$ & $26.55\pm0.55$ \\
231.6978 & $24.98\pm0.15$ & $25.76\pm0.31$ & $25.75\pm0.30$ & $25.12\pm0.17$ & $24.75\pm0.12$ \\
231.7328 & $25.01\pm0.19$ & $25.30\pm0.21$ & $25.81\pm0.32$ & $25.32\pm0.21$ & $24.73\pm0.12$ \\
231.7994 & $25.74\pm0.28$ & $25.41\pm0.21$ & $25.61\pm0.25$ & $24.59\pm0.10$ & $25.10\pm0.16$ \\
232.6975 & $25.39\pm0.22$ & $25.77\pm0.31$ & $25.11\pm0.17$ & $24.91\pm0.14$ &     . . .      \\
232.7382 & $25.13\pm0.26$ &     . . .      & $25.32\pm0.31$ & $24.44\pm0.14$ & $24.92\pm0.23$ \\
233.6447 & $24.95\pm0.18$ & $25.52\pm0.32$ & $24.75\pm0.15$ & $25.14\pm0.22$ &     . . .      \\
233.6767 &     . . .      &     . . .      & $24.63\pm0.17$ & $24.32\pm0.13$ & $25.58\pm0.41$ \\
233.7161 &     . . .      & $25.03\pm0.22$ & $24.92\pm0.19$ &     . . .      &     . . .      \\
233.7628 &     . . .      &     . . .      & $24.88\pm0.23$ & $24.93\pm0.24$ & $25.27\pm0.33$ \\
233.8578 & $25.62\pm0.32$ & $25.44\pm0.28$ & $25.38\pm0.25$ & $25.01\pm0.18$ & $25.00\pm0.18$ \\
233.9099 & $24.37\pm0.12$ & $24.72\pm0.17$ & $25.63\pm0.37$ & $24.86\pm0.19$ & $25.27\pm0.27$ \\
\tableline
HJD$^a$  & r              & r              & r              & r              & r              \\
230.6435 & $25.30\pm0.23$ & $25.59\pm0.30$ &     . . .      &     . . .      &     . . .      \\
230.8403 & $25.10\pm0.17$ & $25.00\pm0.16$ &     . . .      &     . . .      &     . . .      \\
231.7182 & $25.04\pm0.23$ &     . . .      &     . . .      &     . . .      &     . . .      \\
231.7936 &     . . .      &     . . .      &     . . .      &     . . .      &     . . .      \\
233.7547 & $24.77\pm0.21$ & $25.38\pm0.37$ &     . . .      &     . . .      &     . . .      \\
233.7780 & $24.96\pm0.23$ & $25.80\pm0.49$ &     . . .      &     . . .      &     . . .      \\
\end{tabular}
\end{center}
\end{table}
\clearpage
\begin{table}
\begin{center}
Table \thetable---Continued \\
\begin{tabular}{cccccc}
\\
\tableline
\tableline
        & V31            & V32            & V33            & V34            & V35             \\
\tableline
HJD$^a$  & g              & g              & g              & g              & g              \\
230.6523 & $23.75\pm0.05$ & $24.84\pm0.13$ & $24.45\pm0.10$ & $24.89\pm0.13$ & $25.24\pm0.18$ \\
230.7706 & $23.71\pm0.04$ &     . . .      & $24.25\pm0.06$ & $24.99\pm0.12$ & $25.49\pm0.19$ \\
230.8485 & $23.68\pm0.03$ & $24.54\pm0.07$ & $24.25\pm0.05$ & $24.50\pm0.07$ & $24.56\pm0.07$ \\
230.9188 & $23.76\pm0.04$ & $24.21\pm0.07$ & $24.30\pm0.09$ & $24.55\pm0.09$ & $24.76\pm0.11$ \\
231.6445 & $23.05\pm0.02$ & $24.61\pm0.09$ & $24.32\pm0.07$ & $24.61\pm0.09$ & $25.18\pm0.15$ \\
231.6978 & $23.12\pm0.03$ & $24.93\pm0.14$ & $24.39\pm0.09$ & $25.18\pm0.18$ & $25.28\pm0.20$ \\
231.7328 & $23.03\pm0.03$ & $24.88\pm0.14$ & $24.26\pm0.08$ & $24.93\pm0.15$ & $25.16\pm0.18$ \\
231.7994 & $22.96\pm0.02$ & $24.88\pm0.13$ & $24.14\pm0.07$ & $24.64\pm0.10$ & $26.02\pm0.36$ \\
232.6975 & $23.01\pm0.03$ & $23.97\pm0.06$ & $24.21\pm0.07$ & $24.95\pm0.15$ & $24.72\pm0.12$ \\
232.7382 & $23.02\pm0.04$ & $24.29\pm0.12$ & $24.51\pm0.15$ & $24.71\pm0.18$ & $24.55\pm0.15$ \\
233.6447 & $23.30\pm0.04$ & $24.90\pm0.18$ & $24.52\pm0.13$ & $24.90\pm0.18$ & $25.59\pm0.33$ \\
233.6767 & $23.35\pm0.05$ & $25.38\pm0.35$ & $24.55\pm0.16$ & $24.89\pm0.22$ & $25.58\pm0.41$ \\
233.7161 & $23.29\pm0.04$ & $25.02\pm0.21$ & $24.79\pm0.17$ & $24.76\pm0.17$ & $25.19\pm0.25$ \\
233.7628 & $23.29\pm0.05$ & $25.04\pm0.26$ & $24.67\pm0.19$ & $24.41\pm0.15$ & $25.84\pm0.55$ \\
233.8578 & $23.46\pm0.04$ & $24.62\pm0.13$ & $24.52\pm0.12$ & $24.40\pm0.10$ &     . . .      \\
233.9099 & $23.47\pm0.05$ & $24.35\pm0.12$ & $24.86\pm0.19$ & $24.64\pm0.15$ & $25.54\pm0.35$ \\
\tableline
HJD$^a$  & r              & r              & r              & r              & r              \\
230.6435 & $23.79\pm0.06$ &     . . .      & $23.73\pm0.06$ & $24.82\pm0.15$ & $25.12\pm0.20$ \\
230.8403 & $23.53\pm0.04$ &     . . .      & $23.67\pm0.05$ & $24.73\pm0.12$ & $24.99\pm0.15$ \\
231.7182 & $23.08\pm0.04$ &     . . .      & $23.58\pm0.06$ & $24.92\pm0.21$ & $25.15\pm0.26$ \\
231.7936 & $23.17\pm0.08$ &     . . .      & $23.56\pm0.12$ & $24.64\pm0.31$ & $25.09\pm0.47$ \\
233.7547 & $23.23\pm0.05$ &     . . .      & $23.47\pm0.07$ & $24.96\pm0.25$ & $25.35\pm0.36$ \\
233.7780 & $23.24\pm0.05$ &     . . .      & $23.67\pm0.07$ & $24.50\pm0.15$ &     . . .      \\
\end{tabular}
\end{center}
\end{table}
\clearpage
\begin{table}
\begin{center}
Table \thetable---Continued \\
\begin{tabular}{ccc}
\\
\tableline
\tableline
        & V36            & V37             \\
\tableline
HJD$^a$  & g              & g              \\
230.6523 & $25.24\pm0.18$ & $22.36\pm0.01$ \\
230.7706 & $25.62\pm0.22$ & $22.25\pm0.01$ \\
230.8485 & $24.46\pm0.07$ & $22.30\pm0.01$ \\
230.9188 & $24.71\pm0.11$ & $22.34\pm0.01$ \\
231.6445 & $25.07\pm0.14$ & $22.43\pm0.01$ \\
231.6978 & $25.49\pm0.24$ & $22.47\pm0.02$ \\
231.7328 & $25.42\pm0.23$ & $22.47\pm0.02$ \\
231.7994 & $25.35\pm0.20$ & $22.50\pm0.01$ \\
232.6975 & $24.90\pm0.14$ & $22.60\pm0.02$ \\
232.7382 & $24.70\pm0.18$ & $22.65\pm0.03$ \\
233.6447 & $24.34\pm0.14$ & $22.79\pm0.03$ \\
233.6767 & $24.71\pm0.19$ & $22.82\pm0.03$ \\
233.7161 & $24.45\pm0.12$ & $22.81\pm0.03$ \\
233.7628 & $25.02\pm0.26$ & $22.77\pm0.03$ \\
233.8578 & $25.78\pm0.37$ & $22.82\pm0.03$ \\
233.9099 & $25.27\pm0.27$ & $22.90\pm0.03$ \\
\tableline
HJD$^a$  & r              & r              \\
230.6435 & $24.99\pm0.18$ & $22.05\pm0.01$ \\
230.8403 & $24.59\pm0.11$ & $22.02\pm0.01$ \\
231.7182 & $25.15\pm0.25$ & $22.14\pm0.02$ \\
231.7936 & $24.86\pm0.38$ & $22.21\pm0.03$ \\
233.7547 & $24.68\pm0.19$ & $22.43\pm0.03$ \\
233.7780 & $24.73\pm0.18$ & $22.40\pm0.02$ \\
\end{tabular}
\end{center}
\end{table}

\clearpage
\begin{deluxetable}{rrrccccl}
\tablecaption{PvdB Variables. \label{tab_pvdb}}
\tablewidth{0pt}

\tablehead{
\colhead{ID\tablenotemark{a}} &
\colhead{X\tablenotemark{b}} &
\colhead{Y\tablenotemark{b}} &
\colhead{B\tablenotemark{a}} &
\colhead{g\tablenotemark{b}} &
\colhead{r\tablenotemark{b}} &
\colhead{Image\tablenotemark{c}} &
\colhead{Notes\tablenotemark{d}}}
\startdata
 1 & 1173.70 & 1576.26 & $25.41 \pm 0.04$ &       . . .      &       . . .      & extended & \\
 2 & 1931.27 & 1380.25 & $26.22 \pm 0.14$ & $25.66 \pm 0.13$ & $25.76 \pm 0.21$ & blended  & $\chi^2 = 0.83$ \\
 3 & 1680.58 & 1681.29 & $25.53 \pm 0.03$ & $25.16 \pm 0.08$ & $25.21 \pm 0.16$ & blended  & $\chi^2 = 1.83$ \\
 4 & 1396.36 & 1771.24 & $25.80 \pm 0.05$ & $25.03 \pm 0.08$ & $24.24 \pm 0.06$ & blended  & $\chi^2 = 1.01$ \\
 5 & 1781.41 & 1309.26 & $25.82 \pm 0.06$ &       . . .      &       . . .      & extended & \\
 6 & 1608.40 & 1683.98 & $25.91 \pm 0.05$ & $24.84 \pm 0.06$ & $24.35 \pm 0.06$ & good     & $\chi^2 = 0.86$ \\
 7 & 1147.99 & 1934.58 & $25.04 \pm 0.03$ & $24.55 \pm 0.11$ & $24.25 \pm 0.15$ & blended  & $\chi^2 = 5.3$, V08 \\
 8 & 1353.63 & 1472.24 & $25.26 \pm 0.09$ & $24.60 \pm 0.06$ & $23.74 \pm 0.04$ & blended  & $\chi^2 = 1.34$ \\
 9 & 1232.55 & 1709.61 & $25.95 \pm 0.06$ &       . . .      &       . . .      & extended & \\
10 & 1404.24 & 1622.33 & $25.61 \pm 0.05$ & $25.32 \pm 0.10$ & $24.74 \pm 0.08$ & blended  & $\chi^2 = 1.32$ \\
11 & 1183.98 & 1825.01 & $25.60 \pm 0.04$ &       . . .      &       . . .      & extended & \\
12 & 1162.22 & 1730.85 & $26.15 \pm 0.09$ &       . . .      &       . . .      & extended & \\
13 & 1508.68 & 1504.56 & $25.98 \pm 0.07$ & $25.16 \pm 0.09$ & $24.62 \pm 0.07$ & good     & $\chi^2 = 1.02$ \\
14 & 1942.63 & 1326.57 & $25.53 \pm 0.04$ &       . . .      &       . . .      & blended  & \\
15 & 1974.12 & 1209.61 & $25.21 \pm 0.08$ & $24.63 \pm 0.07$ & $24.65 \pm 0.10$ & good     & $\chi^2 = 2.9$, V21 \\
16 & 2001.53 & 1865.98 & $25.38 \pm 0.04$ &       . . .      &       . . .      & extended & \\
17 & 1539.16 & 1699.02 & $25.73 \pm 0.08$ & $25.13 \pm 0.15$ & $25.28 \pm 0.15$ & good     & $\chi^2 = 4.4$, V13 \\
18 & 1036.51 & 2184.48 & $25.68 \pm 0.05$ & $25.40 \pm 0.11$ & $25.16 \pm 0.11$ & good     & $\chi^2 = 1.23$ \\
19 & 1046.94 & 2090.75 & $26.07 \pm 0.07$ &       . . .      &       . . .      & blended  & \\
20 &  966.82 & 1782.99 & $24.66 \pm 0.03$ & $23.82 \pm 0.03$ & $23.19 \pm 0.02$ & blended  & $\chi^2 = 1.45$ \\
21 &  797.42 & 1438.77 & $25.73 \pm 0.06$ & $24.76 \pm 0.05$ & $24.57 \pm 0.06$ & good     & $\chi^2 = 1.74$ \\
22 & 1615.84 & 1745.83 & $25.73 \pm 0.06$ &       . . .      &       . . .      & extended & \\
23 & 2025.98 & 1274.06 & $25.91 \pm 0.05$ &       . . .      &       . . .      & extended & \\
24 & 1327.23 & 2155.01 & $25.64 \pm 0.05$ &       . . .      &       . . .      & extended & \\
25 & 1534.68 & 1562.20 & $26.21 \pm 0.04$ & $25.55 \pm 0.11$ &       . . .      & blended  & $\chi^2 = 0.68$ \\
26 & 1588.10 & 2052.74 & $25.67 \pm 0.05$ & $24.58 \pm 0.06$ & $23.25 \pm 0.02$ & good     & $\chi^2 = 1.02$ \\
27 & 1296.68 & 1668.02 & $25.07 \pm 0.03$ & $24.10 \pm 0.04$ & $23.37 \pm 0.02$ & blended  & $\chi^2 = 1.63$ \\
28 &  966.37 & 1511.18 & $25.18 \pm 0.05$ & $24.70 \pm 0.06$ & $24.37 \pm 0.07$ & good     & $\chi^2 = 1.51$ \\
29 &  894.19 & 1650.01 & $25.91 \pm 0.05$ &       . . .      &       . . .      & extended & \\
30 & 1227.42 & 1903.87 & $25.57 \pm 0.05$ &       . . .      &       . . .      & extended & \\
31 & 1117.94 & 2051.44 & $25.03 \pm 0.03$ &       . . .      &       . . .      & blended  & \\
32 & 1029.00 & 1881.00 & $25.39 \pm 0.04$ &       . . .      & $23.86 \pm 0.05$ & bad col  & \\

\enddata
\tablenotetext{a}{ID numbers and mean B magnitudes from PvdB}
\tablenotetext{b}{Star positions and $g$ and $r$ magnitudes taken from our template images.  For stars 7, 15, and 17 (our V08, V21, and V13), our mean magnitudes are given in place of the template image magnitudes.}
\tablenotetext{c}{Object classification based on our best-seeing images.  Note that some blends could be photometered accurately on our images but could not have been on the PvdB images; for these and the good detections we give $g$ and $r$ magnitudes.}
\tablenotetext{d}{For stars we could photometer accurately, we list the $\chi^2$ of the $g$ photometry (a test of variability) and, if applicable, our variable star identification number.}
\end{deluxetable}

\end{document}